\documentclass[aps,pra,twocolumn, superscriptaddress, showpacs]{revtex4-1}
\usepackage{hyperref}
\usepackage{amsmath}
\usepackage{amssymb,amsfonts}
\usepackage{graphicx}
\usepackage{relsize}
\usepackage{lmodern}
\usepackage{scalefnt}
\usepackage{subfigure}
\usepackage{xspace} 
\usepackage{ifpdf}
\usepackage{pgffor}
\usepackage{array} 
\usepackage{stackrel}
\newcolumntype{C}{>{$}c<{$}} 
\usepackage{tikz}
\usetikzlibrary{shapes.geometric}
\usepackage{color}
\usetikzlibrary{arrows,matrix,calc,scopes,decorations.markings}
\allowdisplaybreaks[1]

\newcommand{\ket}[1]{|{#1}\rangle}
\newcommand{\bra}[1]{\langle{#1}|}

\makeatletter
\newcommand*{\Relbarfill@}{\arrowfill@\Relbar\Relbar\Relbar}
\newcommand*{\xeq}[2][]{\ext@arrow 0055\Relbarfill@{#1}{#2}}
\makeatother
\newcommand{\be}{ \begin{equation}}
\newcommand{\ee}{\end{equation}}
\newcommand{\QFT}{\mbox{\tiny QFT}}
\newcommand{\UV}{\mbox{\tiny UV}}

\begin{document}

\title{Continuous entanglement renormalization on the circle }

\author{Ling-Yan Hung}
\email{lyhung@fudan.edu.cn}
\affiliation{State Key Laboratory of Surface Physics, Fudan University, 200433 Shanghai, China}
\affiliation{Department of Physics and Center for Field Theory and Particle Physics, Fudan University, Shanghai 200433, China}
\affiliation{Institute for Nanoelectronic devices and Quantum computing, Fudan University, 200433 Shanghai, China}
\author{Guifr\'e Vidal}
\affiliation{Sandbox@Alphabet, Mountain View, CA 94043, USA} \affiliation{Perimeter Institute for Theoretical Physics, 31 Caroline St. N., Waterloo, Ontario, Canada}
\date{\today}
    

\begin{abstract}
The continuous multi-scale entanglement renormalization ansatz (cMERA) is a variational class of states for quantum fields.  As originally formulated, the cMERA applies to infinite systems only. In this paper we generalize the cMERA formalism to a finite circle, which we achieve by wrapping the action of the so-called \textit{entangler} around the circle. This allows us to transform a cMERA on the line into a cMERA on the circle. In addition, in the case of a Gaussian cMERA for non-interacting quantum fields, the method of images allow us to prove the following result: if on the line a cMERA state is a good approximation to a ground state of a local QFT Hamiltonian, then (under mild assumptions on their correlation functions) the resulting cMERA on a circle is also a good approximation to the ground state of the same local QFT Hamiltonian on the circle. 
\end{abstract}
\maketitle

The multi-scale entanglement renormalization ansatz (MERA) \cite{MERA1,MERA2} is a tensor network used to efficiently represent a class of quantum many-body wavefunctions on the lattice and as the basis for non-perturbative, numerical approaches to \textit{strongly interacting} quantum systems \cite{MERA1,MERA2,MERA3,MERA4,MERA5,MERA6,MERA7,MERA8}. In contrast to quantum monte carlo methods, MERA algorithms are not affected by the so-called sign problem, and can therefore be applied to e.g. systems of interacting fermions and frustrated quantum magnets. Other applications include statistical mechanics \cite{MERA_stat}, quantum gravity \cite{MERA_QG1,MERA_QG2,MERA_QG3,MERA_QG4,MERA_QG5}, cosmology \cite{MERA_cosmo1,MERA_cosmo2,MERA_cosmo3}, error correction \cite{MERA_EC1,MERA_EC2} or machine learning \cite{MERA_ML1, MERA_ML2, MERA_ML3,MERA_ML4}.

In pioneering work \cite{cMERA}, Haegeman, Osborne, Verschelde and Verstraete proposed the continuous MERA (cMERA) for quantum field theories (QFTs), a generalization of MERA from the lattice to the continuum. These authors also showed, with explicit examples, that cMERA can accurately approximate the long-distance properties of ground states of \textit{non-interacting} QFTs. The ultimate goal of the cMERA program is to become a non-perturbative approach to \textit{strongly interacting} QFTs (see \cite{cMERA_int1,cMERA_int2,cMERA_int3,cMERA_int4} for steps in this direction), and in this way mirror the success of MERA on the lattice. In the meantime, however, the \textit{Gaussian} cMERA for non-interacting QFTs \cite{cMERA,Qi,Adrian,Yijian,Adrian2} has already proven its worth. Besides providing a valuable proof of concept that the lattice MERA formalism can be carried over to the continuum, Gaussian cMERA has been used as a toy model also in the context of the AdS/CFT correspondence and quantum gravity \cite{cMERAholo1,cMERAholo2,cMERAholo3,cMERAholo4,cMERAholo5,cMERAholo6,cMERAholo7,cMERAholo8,cMERAholo8}. Importantly, the Gaussian cMERA recasts key aspects of lattice MERA ---originally formulated in terms of networks of tensors and numerical optimizations--- in the language of free quantum fields, which is familiar to a broader spectrum of theoretical physicists. For instance, the interpretation of the lattice ansatz as an entangling evolution in scale (as revisited below), or its connection to conformal data at a quantum critical point \cite{Qi} become very clear in their continuum, free field realizations. In this sense, Gaussian cMERA contributes significantly to the conceptual foundations of the MERA formalism.


In this work we generalize the cMERA to finite systems. Specifically, given a cMERA on the infinite line, we explain how to build a cMERA on a finite circle. This is accomplished by modifying the action of the cMERA's \textit{entangler} on the line so that it wraps around the circle. Importantly, in the Gaussian case, where our proposal amounts to using the \textit{method of images} [A3], we will show the following result. If the initial cMERA on the line is a good approximation to the ground state of a QFT, then (under simple assumptions on the correlation functions) the resulting cMERA on the circle is guaranteed to also be a good approximation to the ground state of the QFT on that circle. We illustrate our findings with a relativistic free boson in one dimension.
However, these results can also be generalized to higher dimensions, to build e.g. a cMERA on a cylinder or a torus [A4]. A related generalization to systems with open boundary conditions and/or defects has been recently proposed by Franco-Rubio \cite{Adrian3}. 

\textit{cMERA on the line.---} Let us start by considering a bosonic quantum field on the line. It is characterized by field operators $\phi(x)$, $\pi(x)$ with $x\in\mathbb{R}$ and canonical commutation relations $[\phi(x),\pi(y)] = i \delta(x-y)$. We denote their Fourier transforms by $\phi(k)$, $\pi(k)$ [A1].
Following the original proposal of Ref. \cite{cMERA}, a one-parameter family of cMERA states $\ket{\Psi^{\Lambda}(s)}$ is then defined as 
\begin{equation} \label{eq:cMERA_line}
    \ket{\Psi^{\Lambda}(s)} \equiv \exp \left(- i s\left(L+K \right) \right) \ket{\Lambda}.
\end{equation}
Here $\ket{\Lambda}$ is an unentangled state characterized by
\begin{equation} \label{eq:Lambda}
    \left( \sqrt{\frac{\Lambda}{2}} \phi(k) + \frac{i}{\sqrt{2\Lambda}} \pi(k)\right) \ket{\Lambda} = 0,~~~~ \forall k \in \mathbb{R},
\end{equation}
$L$ is the non-relativistic re-scaling operator 
\begin{eqnarray}
   L \equiv \frac{1}{2}\int_{\mathbb{R}} dk [\pi(-k) (k \partial_k + 1/2)\phi(k) + h.c], 
\end{eqnarray}
and $K$ is the quasi-local \textit{entangler}, given by \cite{Kconstant}
\begin{eqnarray} \label{eq:K}
K &\equiv& \frac{1}{2}\int_{\mathbb{R}_{2}} dxdy ~g(x-y) [\pi(x)\phi(y) + \phi(x)\pi(y)]~~~\\
&=&\frac{1}{2}\int_{\mathbb{R}} dk\, g(k) [\pi(-k)\phi(k) + \phi(-k)\pi(k)],
\end{eqnarray}
where the entangling profile $g(x)$ is picked at $x=0$ and vanishes \textit{exponentially} fast for large $|x| \gg 1/\Lambda$, with $1/\Lambda$ a characteristic length scale, (e.g. $g(x) \sim e^{-\Lambda|x|}$ in Ref. \cite{Yijian}) and $g(k)\equiv \int_{\mathbb{R}}\!dx~e^{-ikx} g(x)$ is its Fourier transform. 

\begin{figure}
	\centering
	\includegraphics[width=1.0\linewidth]{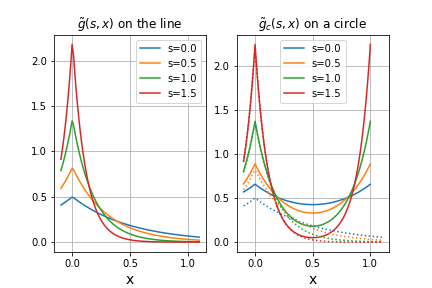}
	\caption{(Left) Example of entangling profile $\tilde{g}(s,x)$ on the line , which is picked at $x=0$ and has width $e^{-s}/\Lambda$ that decays exponentially with the scale parameter $s$.
	(Right) Corresponding entangling profile $\tilde{g}_c(s,x)$ for a circle of size $l_c=1$, which is periodic by construction. These examples correspond to Eqs. \eqref{eq:gstar} and \eqref{eq:summed} (excluding $-\delta(x)/2$), for $\Lambda=m=2$.}
	\label{fig:ggc}
\end{figure}

\textit{Re-scaled picture.---} In this work we use a different picture re-scaled by the relativistic re-scaling operator 
\begin{eqnarray}
   D \equiv \frac{1}{2}\int_{\mathbb{R}} dk [\pi(-k) (k \partial_k + 1)\phi(k) + h.c]. 
\end{eqnarray}
The cMERA state is then $\ket{\tilde{\Psi}^{\Lambda}(s)} \equiv e^{isD} \ket{\Psi^{\Lambda}(s)}$, or
\begin{equation} \label{eq:cMERAtilde_line}
    \ket{\tilde{\Psi}^{\Lambda}(s)}  = \mathcal{P}\exp{\left(-i \int_0^{s} ds'~\tilde{K}(s')\right)} \ket{\Lambda},
\end{equation}
where $\mathcal{P}\exp$ is a path ordered exponential and $\tilde{K}(s) \equiv e^{isD} (L+K-D) e^{-isD}$ is a scale-dependent entangler with
\begin{eqnarray}
 \tilde{K}(s) &=& \frac{1}{2}\int_{\mathbb{R}_{2}}\!\!dxdy ~\tilde{g}(s,x-y)[\pi(x)\phi(y) + \phi(x)\pi(y)]~~~~~\\
&=&\frac{1}{2}\int_{\mathbb{R}} dk\, \tilde{g}(s,k) [\pi(-k)\phi(k) + \phi(-k)\pi(k)].
\end{eqnarray}
with $\tilde{g}(s,x) \equiv e^{s}g(e^{s}x) - \delta(x)/2$ and $\tilde{g}(s,k) = g (e^{-s}k)-1/2$. Eq. \eqref{eq:cMERAtilde_line} describes a so-called \textit{entangling evolution in scale}.  Starting from the unentangled state $\ket{\Lambda}$, we act with the entangler $\tilde{K}(s)$, which initially (that is, for $s=0$) introduces entanglement at scale $1/\Lambda$. As the evolution progresses, the entangler's characteristic length decreases exponentially with growing $s$ as $e^{-s}/\Lambda$, see Fig. \ref{fig:ggc}, in such a way that entanglement is introduced at shorter and shorter distances. The resulting cMERA state $\ket{\tilde{\Psi}^{\Lambda}(s)}$ is thus entangled in the range of distances $[e^{-s}/\Lambda, 1/\Lambda]$ \cite{cutoff}. Qualitatively, this entanglement structure is compatible with the ground state of a massive QFT with mass $m\sim \Lambda$ (or correlation length $1/\Lambda$) and a UV cut-off at momentum $k \sim e^{s}\Lambda$ (or UV length $e^{-s}/\Lambda$).


\begin{figure}
\centering
\includegraphics[width=1.0\linewidth]{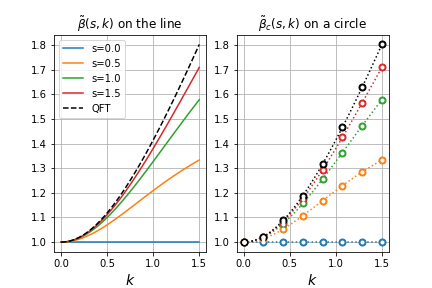}
    \caption{[beta.png] (Left) Function $\tilde{\beta}(s,k)$ characterizing the operators in Eqs. \eqref{eq:bLamskPsi}-\eqref{eq:bLamsk} that annihilate the cMERA state $\ket{\tilde{\Psi}^{\Lambda}(s)}$ on the line, for the specific choice in Eq. \eqref{eq:beta_star_line}. For $s=0$ it produces the constant $\Lambda$ (corresponding to the unentangled state $\ket{\Lambda}$), whereas for large $s$ it tends to $\beta^{\QFT}(k)$ in Eq. \eqref{eq:beta_QFT_line} for a relativistic boson with mass $m=\Lambda=1$. 
    (Right)	On a circle of length $l_c$, the method of images leads to a closely related function $\tilde{\beta}_c(s,n)$ that samples $\tilde{\beta}(s,k)$ at discrete momenta $k_n = 2\pi n /l_c$, see Eq. \eqref{eq:beta_circle}. }
	\label{fig:beta}
\end{figure}

\textit{Annihilation operators and correlation functions.---} Since the above cMERA is the result of evolving the state $\ket{\Lambda}$, which is Gaussian, by the entangler $\tilde{K}(s)$, which is quadratic, it follows that $\ket{\tilde{\Psi}^{\Lambda}(s)}$ is also Gaussian and can thus be characterized in terms of some complete set of annihilation operators $\tilde{b}^{\Lambda}(s,k)$, that is 
\begin{equation} \label{eq:bLamskPsi}
    \tilde{b}^{\Lambda}(s,k) ~\ket{\tilde{\Psi}^{\Lambda}(s)} = 0,~~~\forall k \in \mathbb{R}
\end{equation}
where we make the following ansatz
\begin{equation} 
\tilde{b}^{\Lambda}(s,k) \equiv \sqrt{\frac{\tilde{\beta}(s,k)}{2}} \phi(k) + \frac{1}{\sqrt{2\tilde{\beta}(s,k)}} \pi(k). \label{eq:bLamsk}
\end{equation}
By imposing that $b^{\Lambda}(s,k)$ be the result of an entangling evolution in scale by $\tilde{K}(s)$ (namely, $\frac{\partial}{\partial s} \tilde{b}^{\Lambda}(s,k) = -i [\tilde{K}(s), \tilde{b}^{\Lambda}(s,k)]$), we obtain the differential equation
\begin{equation} \label{eq:diff_line}
    \frac{\partial}{\partial s} \tilde{\beta}(s,k) = -2\tilde{g}(s,k) \tilde{\beta}(s,k),~~~~\forall k \in \mathbb{R},
\end{equation}
relating $\tilde{g}(s,k)$ and $\tilde{\beta}(s,k)$, complemented by the initial condition $\tilde{\beta}(0,k) =\Lambda$, which enforces that the unentangled state $\ket{\tilde{\Psi}^{\Lambda}(0)} = \ket{\Lambda}$ is recovered at $s=0$, see Fig. \ref{fig:beta}. Finally, from Eq. \eqref{eq:bLamskPsi} one obtains two-point correlation functions $C_{AB}(s, x-y) \equiv \bra{\tilde{\Psi}^{\Lambda}(s)} A(x)B(y)\ket{\tilde{\Psi}^{\Lambda}(s)} $ [A2],
\begin{eqnarray} \label{eq:corr1}
C_{\phi\phi}(s, x) &=& \int_{\mathbb{R}} \frac{dk}{2\pi} ~  e^{i kx}\frac{1}{2\tilde{\beta}(s,k)},~~~\\
C_{\pi\pi}(s, x) &=& \int_{\mathbb{R}} \frac{dk}{2\pi} ~  e^{i kx}\frac{\tilde{\beta}(s,k)}{2}.~~~~~ \label{eq:corr2}
\end{eqnarray}
In summary, given an entangling profile $g(x)$ on the line, the resulting cMERA (including its correlation functions) is characterized by the solution $\tilde{\beta}(s,k)$ of Eq. \eqref{eq:diff_line}. Next we generalize this strategy to the circle.

\textit{cMERA on the circle.---} Consider a bosonic field on a circle of size $l_c$, with field operators $\phi(x),\pi(x)$ for $x\in[0,l_c)$ and $[\phi(x),\pi(y)]=i\delta(x-y)$. Their Fourier modes
\begin{eqnarray}
\phi(n) &\equiv& \frac{1}{\sqrt{l_c}} \int_0^{l_c} dx ~e^{-i k_{n}x} \phi(x), \\
\pi(n) &\equiv&  \frac{1}{\sqrt{l_c}} \int_0^{l_c} dx ~e^{-i k_{n}x}  \pi(x),
\end{eqnarray}
are now discrete, with $n \in \mathbb{Z}$ and $k_n \equiv 2\pi n/l_c$.
By analogy with \eqref{eq:cMERAtilde_line}, we define the cMERA state 
\begin{equation} \label{eq:cMERAtilde_circle}
    \ket{\tilde{\Psi}_c^{\Lambda}(s)} \equiv \mathcal{P}\exp{\left(-i \int_0^{s} ds'~\tilde{K}_c(s')\right)} \ket{\Lambda_c}.
\end{equation}
Here $\ket{\Lambda_c}$ is the unentangled state characterized by
\begin{equation} \label{eq:Lambda_c}
    \left( \sqrt{\frac{\Lambda}{2}} \phi(n) + \frac{i}{\sqrt{2\Lambda}} \pi(n)\right) \ket{\Lambda_c} = 0,~~~~ \forall n \in \mathbb{Z},
\end{equation}
and the scale-dependent entangler $\tilde{K}_c(s)$ is given by
\begin{eqnarray}
 \tilde{K}_c(s) &\equiv& \frac{1}{2}\!\int_{0}^{l_c}\!\!\! dxdy ~\tilde{g}_c(s,x\!-\!y)[\pi(x)\phi(y)\! + \!\phi(x)\pi(y)]~~~~\\
&=& \frac{1}{2} \sum_{n\in \mathbb{N}} \tilde{g}_c(s,n) [\pi(-n)\phi(n) + \phi(-n)\pi(n)],
\end{eqnarray}
where $\tilde{g}_c(s,x)$ is some profile function and $\tilde{g}_c(s,n) \equiv \int_{0}^{l_c} dx~e^{-ik_{n}x} \tilde{g}_c(x)$ is its discrete Fourier transform. 

Our key proposal is to build the profile function $\tilde{g}_c(s,x)$ on the circle from the profile function $\tilde{g}(s,x)$ on the line by adding contributions that come from wrapping the latter around the circle,
\begin{equation} \label{eq:gc}
    \tilde{g}_c(s,x) \equiv \sum_{n\in \mathbb{Z}} \tilde{g}(s,x+nl_c),
\end{equation}
see Fig. \ref{fig:ggc}. 
Eq. \eqref{eq:gc} is seen to imply, through simple but tedious calculations [A2,A3], that most properties of the entangler and cMERA on the circle are closely related to those of the original constructions on the line.  For starters, the Fourier transform $\tilde{g}_c(s,n)$ of $\tilde{g}_c(s,x)$ in \eqref{eq:gc} is given simply by [A3]
\begin{equation} \label{eq:gcn}
    \tilde{g}_c(s,n) =  \tilde{g}(s,k_n),
\end{equation}
that is, in terms of the function $\tilde{g}(s,k)$ on the line evaluated at discrete values $k=k_n$ of the momentum. The resulting Gaussian cMERA  $\ket{\tilde{\Psi}_c^{\Lambda}(s)}$ is again characterized by a set of annihilation operators $b^{\Lambda}(s,n)$,
\begin{eqnarray} \label{eq:bLamPsi_circle}
 &&\tilde{b}^{\Lambda}(s,n) ~\ket{\tilde{\Psi}_c^{\Lambda}(s)} = 0,~~~\forall n \in \mathbb{Z},\\
 &&\tilde{b}^{\Lambda}(s,n) \equiv \sqrt{\frac{\tilde{\beta}_c(s,n)}{2}} \phi(n) + \frac{1}{\sqrt{2\tilde{\beta}_c(s,n)}} \pi(n), \label{eq:bLam_circle}
\end{eqnarray}
where coefficients $\tilde{\beta}_c(s,n)$ are determined by imposing $\frac{\partial}{\partial s} \tilde{b}^{\Lambda}(s,n) = -i [\tilde{K}_c(s), \tilde{b}^{\Lambda}(s,n)]$. This results in
\begin{equation} \label{eq:diff_circle}
    \frac{\partial}{\partial s} \tilde{\beta}_c(s,n) = -2\tilde{g}_c(s,n) \tilde{\beta}_c(s,n),~~~~\forall n \in \mathbb{Z},
\end{equation}
which is identical to Eq. \eqref{eq:diff_line} for $k=k_n$. Thus $\tilde{\beta}_c(s,n)$ on the circle is also given in terms of $\tilde{\beta}(s,k)$ on the line,
\begin{equation} \label{eq:beta_circle}
    \tilde{\beta}_c(s,n) = \tilde{\beta}(s,k_n),
\end{equation}
see Fig. \eqref{fig:beta}. Finally, from Eq \eqref{eq:bLamPsi_circle} we obtain correlation functions $C_{c,AB}(s, x - y) \equiv \bra{\tilde{\Psi}_c^{\Lambda}(s)} A(x)B(y)\ket{\tilde{\Psi}_c^{\Lambda}(s)} $, 
which again relate simply to correlators on the line [A3]: 
\begin{eqnarray} \label{eq:corr1_circle}
C_{c,\phi\phi}(s, x) 
&=& \sum_{n\in \mathbb{Z}}  C_{\phi\phi}(s,x+nl_c),~~~~ \\
C_{c,\pi\pi}(s, x) 
&=&\sum_{n\in \mathbb{Z}} C_{\pi\pi}(s,x+nl_c).\label{eq:corr2_circle}~~~
\end{eqnarray}
Therefore starting from a cMERA on the line, specified in terms of an entangling profile function $g(x)$, we have explained how to produce a closely related cMERA on the circle.

\begin{figure}
	\centering
	\includegraphics[width=1.0\linewidth]{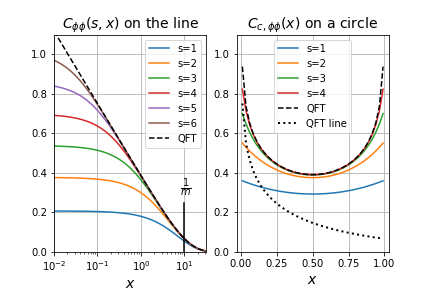}
	\caption{ (Left) Correlation function $C^{\QFT}_{\phi\phi}(x)$ for the QFT ground state on the line with mass $m=0.1$ (discontinuous line) and $C_{\phi\phi}(s,x)$ for the corresponding cMERA with $\Lambda= m$. The cMERA correlator offers an accurate approximation for $x$ larger than $x_{\UV} = e^{-s}/m$.  (Right) Same correlation functions on a circle of length $l_c=1$.}
	\label{fig:corr_2in1}
\end{figure}


Suppose now that the cMERA state $\ket{\tilde{\Psi}^{\Lambda}(s)}$ approximates the ground state $\ket{\Psi^{\QFT}}$ of a non-interacting QFT on the line, in that the relative error $E_{AB}(s,x)$ between their correlators is at most some small $\epsilon(s)>0$, i.e.
\begin{eqnarray}
 E_{AB}(s,x) \equiv \left|\frac{C_{AB}(s, x)- C^{\QFT}_{AB}(x)}{C^{\QFT}_{AB}(x)}\right| \leq \epsilon(s),~~~
\end{eqnarray}
for all $x \geq x_{\UV}$, where $x_{\UV}$ is some UV length. Then, if $C^{\QFT}_{AB}(x)$ does not change sign for $x \geq x_{\UV}$, the proposed cMERA $\ket{\tilde{\Psi}_c^{\Lambda}(s)}$ on the circle also approximates the ground state $\ket{\Psi_c^{\QFT}}$ of the same QFT on the circle, in that, for all $x\geq x_{\UV}$, their correlators also fulfil
\begin{eqnarray}
 E_{c,AB}(s,x) \equiv \left|\frac{C_{c,AB}(s, x)- C^{\QFT}_{c,AB}(x)}{C^{\QFT}_{c,AB}(x)}\right| \leq \epsilon(s).~~~~~~
\end{eqnarray}
This result (see [A2] for a proof) follows quite simply from the fact that the QFT correlators on the line and on the circle obey relations analogous to Eqs. \eqref{eq:corr1_circle}-\eqref{eq:corr2_circle} .

\textit{Example.---} Let  $\ket{\Psi^{\QFT}}$ denote the ground state of the relativistic, free boson Hamiltonian $H^{\QFT} \equiv \int_{\mathbb{R}} dx ~h(x)$ on the line, with local Hamiltonian density
\begin{equation} \label{eq:h}
h(x) \equiv \frac{1}{2}\left[ \pi(x)^2 + \left(\partial_x \phi(x)\right)^2 + m^2\phi(x)^2\right],
\end{equation}
where $m$ is the mass. As reviewed in [A1], $\ket{\Psi^{\QFT}}$ is annihilated by operators $b(k)$, i.e. $b(k) \ket{\Psi^{\QFT}}= 0$, with
\begin{eqnarray}
 b(k) &\equiv& \sqrt{\frac{\beta^{\QFT}(k)}{2}} \phi(k) + \frac{1}{\sqrt{2\beta^{\QFT}(k)}} \pi(k), \\
 \beta^{\QFT}(k) &\equiv& \sqrt{k^2 + m^2}.
\end{eqnarray}
This leads to correlation functions such as
\begin{equation}
    C^{\QFT}_{\phi\phi}(x) = K_0(mx) \sim \frac{1}{\sqrt{8\pi}}\frac{e^{-mx}}{\sqrt{mx}},
\end{equation}
where $K_0$ is a modified Bessel function of second kind and the expansion is for large distances $mx \gg 1$. Following \cite{Yijian}, a cMERA approximation $\ket{\tilde{\Psi}^{\Lambda}(s)}$ is obtained with the so-called \textit{magic} entangling profile $g(x) = \frac{\Lambda}{4}e^{-\Lambda |x|}$, which in the rescaled picture leads to
\begin{eqnarray} \label{eq:gstar}
 \tilde{g}(s,x) &=& \frac{e^{s}\Lambda }{4}e^{-e^s\Lambda |x|} -\frac{\delta(x)}{2}, \\
\tilde{g}(s,k) &=& \frac{\Lambda^2}{2(\Lambda^2+e^{2s}k^2)} - \frac{1}{2}.~~~
\end{eqnarray} 
Differential equation \eqref{eq:diff_line} is then solved by
\begin{equation} \label{eq:beta_star_line}
\tilde{\beta}(s,k) =  \frac{\sqrt{k^2 + \Lambda^2}}{\sqrt{1+(e^{-s}k/\Lambda)^2}}, ~~~~~~
\end{equation}
which indeed closely mimics $\beta^{\QFT}(k)$ for $\Lambda=m$, 
\begin{eqnarray}
\tilde{\beta}(s,k)
 &=& \beta^{\QFT}(k) \left(1 + O\left(\left(\frac{e^{-s}k}{m}\right)^2\right) \right), 
\end{eqnarray}
see Fig. \eqref{fig:beta}. As a result, cMERA correlators accurately approximate QFT correlators, see Fig. \eqref{fig:corr_2in1} and [A2]. For instance, for the $\phi\phi$ correlator we find
\begin{equation} \label{eq:Cphiphi_comparison}
    C_{\phi\phi}(s,x) = C^{\QFT}_{\phi\phi}(x) \left(1 + O\left(e^{-2s}\right)\right), ~~~~mx \gg 1,
\end{equation}
so that the relative error between correlation functions is exponentially suppressed with $s$, $E_{\phi\phi}(s,x) = O(e^{-2s})$. Numerical evaluation shows that the error is small already for $x$ larger than the UV length $x_{\UV} \equiv e^{-s}/m$. Similar results are obtained for the $\pi\pi$ correlator.

Moving to a circle of size $l_c$, we consider the ground state $\ket{\Psi^{\QFT}_c}$ of $H^{\QFT}_c = \int_{0}^{l_c} dx~h(x)$, where $h(x)$ is again the Hamiltonian density in Eq. \eqref{eq:h}. Its correlators are obtained by summing over images on the line, e.g. $C_{c,\phi\phi}^{\QFT}(x) = \sum_{n} C_{\phi\phi}^{\QFT}(x+n l_c)$. Similarly, we build the cMERA's entangling profile $\tilde{g}_c(s,x)$ using Eq. \eqref{eq:gc}. After explicitly summing over images we obtain [A2]
\begin{equation} \label{eq:summed}
\tilde{g}_c(s,x)  = \frac{e^{s}m}{4} \frac{\cosh\left(e^{s}m\left(\frac{l_c}{2}-x\right)\right)}{\sinh\left(e^{s}m \frac{l_c}{2}\right)} -\frac{\delta(x)}{2},~~~
\end{equation}
which is shown in Fig. \eqref{fig:ggc}. The correlation functions of the resulting cMERA on the circle obey Eqs. \eqref{eq:corr1_circle}-\eqref{eq:corr2_circle}. In particular, given that $C^{\QFT}_{\phi\phi}(x)$ on the line is positive for all $x>0$ (that, is, it does not change sign), Eq. \eqref{eq:Cphiphi_comparison} implies that the relative error between ground state and cMERA correlation functions on the circle is also exponentially suppressed with $s$ for $mx \gg 1$, namely $E_{c,\phi\phi}(s,x) = O(e^{-2s})$. Numerical evaluation again shows that the error is small already for $x$ larger than the UV length $x_{\UV} \equiv e^{-s}/m$, see Fig. \eqref{fig:corr_2in1}. Similar results are obtained for $\pi \pi$ correlation functions. 

\textit{Discussion.---} In this work we have extended the cMERA formalism from the infinite line to a finite circle, by wrapping the entangling profile $g(x)$ around the circle. This construction can be generalized straightforwardly to higher spatial dimensions, allowing to also define the cMERA on e.g. a cylinder or a torus [A4]. 
Notice that our proposal, discussed here for a Gaussian cMERA, is also valid for an interacting cMERA (with an entangler that includes terms that are e.g. quartic in the fields). However, only in the Gaussian case were we able to additionally show that, if the cMERA is a good approximation to a non-interacting QFT ground state on the line, then this property also holds on the circle. It is tempting to conclude by conjecturing that an equivalent result may still apply for an interacting cMERA. After all, that is known to be the case on the lattice. Indeed, a lattice MERA numerically optimized to resemble the ground state of an infinite spin chain is routinely used to build a MERA approximation for the ground state of a corresponding finite spin chain \cite{MERA7, MERA8}.

\textit{Acknowledgements.---} LYH acknowledges the support of NSFC (Grant
No. 11922502, 11875111) and the Shanghai Municipal Science and Technology Major Project
(Shanghai Grant No.2019SHZDZX01), and Perimeter Institute for hospitality as a part of the Emmy Noether Fellowship programme. GV is a CIFAR fellow in the Quantum Information Science Program and a Distinguish Visiting Research Fellow at Perimeter Institute. 
Sandbox is a team within the Alphabet family of companies, which includes Google, Verily, Waymo, X, and others. Research at Perimeter Institute is supported by the Government of Canada through the Department of Innovation, Science and Economic Development Canada and by the Province of Ontario through the Ministry of Research, Innovation and Science. 


\vspace{0.5cm}
\textbf{Appendix A1: Entangling evolution on the line}
\vspace{0.5cm}

In this Appendix we derive in more detail the equations describing the entangling evolution in scale that generates the cMERA. Most of the material in this Appendix, concerned with the cMERA on the line, either already appeared scattered through previous work, mostly in \cite{cMERA,Qi,Adrian, Yijian}, or could have been derived there with little effort. In particular, the accuracy analysis of the magic cMERA on the line, including short and long distance expansions, would have fitted naturally in \cite{Yijian}. Notice, however, that in contrast with Ref. \cite{Yijian}, in this appendix we work in the rescaled picture that we used in the main text, which is a most natural choice in order to subsequently analyse a circle of constant size $l_c$. 

\vspace{0.5cm}
\begin{center}
    \textit{A. cMERA on the line}
\end{center}
\vspace{0.5cm}

As in the main text, we consider a bosonic quantum field in one spatial dimension, as given by field operators $\phi(x)$ and $\pi(x)$ for $x\in \mathbb{R}$ with the commutation relation $[\phi(x),\pi(y)] = i \delta(x-y)$. We also introduce Fourier transformed fields
\begin{eqnarray}
\phi(k) &\equiv& \frac{1}{\sqrt{2\pi}} \int_{\mathbb{R}} dx ~e^{-ikx} \phi(x), \\
\pi(k) &\equiv& \frac{1}{\sqrt{2\pi}} \int_{\mathbb{R}} dx ~e^{-ikx} \pi(x),
\end{eqnarray}
which obey $\phi(k)^{\dagger} = \phi(-k)$, $\pi(k)^{\dagger} = \pi(-k)$, and $[\phi(k),\pi(q)] = i \delta(k+q)$. In terms of $\phi(k), \pi(k)$, the original fields $\phi(x), \pi(x)$ read
\begin{eqnarray}
\phi(x) &\equiv& \frac{1}{\sqrt{2\pi}} \int_{\mathbb{R}} dk ~e^{ikx} \phi(k), \\
\pi(x) &\equiv& \frac{1}{\sqrt{2\pi}} \int_{\mathbb{R}} dk ~e^{ikx} \pi(k),
\end{eqnarray}
where we have used
\begin{equation}
    \int_{\mathbb{R}}e^{ikx} dk  = 2\pi\delta(x), ~~~ \int_{\mathbb{R}}e^{ikx} dx = 2\pi\delta(k).~~
\end{equation}

Consider an arbitrary entangler
\begin{eqnarray}
 \tilde{K}(s) &=& \frac{1}{2}\int_{\mathbb{R}_{2}}\!\!dxdy ~\tilde{g}(s,x-y)[\pi(x)\phi(y) + \phi(x)\pi(y)]~~~~~~~\\
&=&\frac{1}{2}\int_{\mathbb{R}} dk\, \tilde{g}(s,k) [\pi(-k)\phi(k) + \phi(-k)\pi(k)],
\end{eqnarray}
where $\tilde{g}(s,x)$ is an entangling profile and $\tilde{g}(s,k) \equiv \int_{\mathbb{R}}\!dx~e^{-ikx} g(s,x)$ is its Fourier transform. We define the cMERA state $\ket{\tilde{\Psi}^{\Lambda}(s)}$ as an entangling evolution by $\tilde{K}(s)$,
\begin{equation} \label{eq:cMERA_def1}
    \ket{\tilde{\Psi}^{\Lambda}(s)} \equiv \mathcal{P}\exp \left(-i \int_0^{s} ds' \tilde{K}(s')  \right) \ket{\Lambda},
\end{equation}
starting from the unentangled state $\ket{\Lambda}$, characterized by
\begin{eqnarray}
    &&\tilde{b}^{\Lambda}(0,k) \ket{\Lambda} = 0,~~~~ \forall k \in \mathbb{R}, \\
    &&\tilde{b}^{\Lambda}(0,k) \equiv \sqrt{\frac{\Lambda}{2}} \phi(k) + \frac{i}{\sqrt{2\Lambda}} \pi(k). \label{eq:initial_R1}
\end{eqnarray}
In order to similarly characterize the cMERA state $\ket{\tilde{\Psi}^{\Lambda}(s)}$ in terms of a complete set of annihilation operators $\tilde{b}^{\Lambda}(s,k)$,
\begin{equation}
    \tilde{b}^{\Lambda}(s,k) \ket{\tilde{\Psi}^{\Lambda}(s)} = 0, ~~~\forall k \in \mathbb{R},
\end{equation}
we assume an expression of the form
\begin{equation} \label{eq:ansatz_b}
    \tilde{b}^{\Lambda}(s,k) \equiv \sqrt{\frac{\tilde{\beta}(s,k)}{2}}\phi(k) + \frac{i}{\sqrt{2\tilde{\beta}(s,k)}}\pi(k),
\end{equation}
and require that $\tilde{b}^{\Lambda}(s,k)$ change with $s$ according to the entangling evolution, that is
\begin{equation} \label{eq:evolve_scale1}
    \frac{\partial}{\partial s} \tilde{b}^{\Lambda}(s,k) = -i [\tilde{K}(s), \tilde{b}^{\Lambda}(s,k)],~~~~\forall k \in \mathbb{R},
\end{equation}
with Eq. \eqref{eq:initial_R1} as the initial conditions at $s=0$.

On the one hand we have
\begin{eqnarray}
 &&\frac{\partial}{\partial s} \tilde{b}^{\Lambda}(s,k) \\
 &=& 
  \frac{\partial}{\partial s} \left( \sqrt{\frac{\tilde{\beta}(s,k)}{2}}\phi(k) + \frac{i}{\sqrt{2\tilde{\beta}(s,k)}}\pi(k) \right)\\
 &=& \frac{\partial_s \tilde{\beta}(s,k)}{2 \tilde{\beta}(s,k)}\left( \sqrt{\frac{\tilde{\beta}(s,k)}{2}}\phi(k) - \frac{i}{\sqrt{2\tilde{\beta}(s,k)}}\pi(k) \right). 
\end{eqnarray}
On the other hand the commutator reads
\begin{eqnarray}
 && [\tilde{K}(s), \tilde{b}^{\Lambda}(s,k)]= \frac{1}{2}\int_{\mathbb{R}} dq ~\tilde{g}(s,q) \times\\
&& \left[\pi(-q)\phi(q) + \phi(-q)\pi(q), \sqrt{\frac{\tilde{\beta}(s,k)}{2}}\phi(k) + \frac{i\pi(k)}{\sqrt{2\tilde{\beta}(s,k)}} \right] \nonumber\\
&=& -i\tilde{g}(s,k) \left(\sqrt{\frac{\tilde{\beta}(s,k)}{2}}\phi(k) - \frac{i}{\sqrt{2\tilde{\beta}(s,k)}}\pi(k)\right).
\end{eqnarray}
Replacing these expressions in Eq. \eqref{eq:evolve_scale1} we conclude that $\tilde{\beta}(s,k)$ must obey the differential equation
\begin{equation} \label{eq:diff1}
    \frac{\partial}{\partial s} \tilde{\beta}(s,k) = -2\tilde{g}(s,k) \tilde{\beta}(s,k),~~~~\forall k \in \mathbb{R},
\end{equation}
with initial condition (cf. Eq. \eqref{eq:initial_R1})
\begin{equation} \label{eq:initial_R}
    \tilde{\beta}(s=0,k) = \Lambda,~~~~\forall k \in \mathbb{R}.
\end{equation}

Its correlation functions in momentum space read 
\begin{eqnarray}
 \bra{\tilde{\Psi}^{\Lambda}(s)} \phi(k)\phi(k')\ket{\tilde{\Psi}^{\Lambda}(s)} &\equiv& \delta(k+k') ~ \tilde{C}^{\Lambda}_{\phi\phi}(s,k),\\
 \bra{\tilde{\Psi}^{\Lambda}(s)} \pi(k)\pi(k')\ket{\tilde{\Psi}^{\Lambda}(s)} &\equiv& \delta(k+k') ~ \tilde{C}^{\Lambda}_{\pi\pi}(s,k),
\end{eqnarray}
for
\begin{equation}
    \tilde{C}^{\Lambda}_{\phi\phi}(s,k) = \frac{1}{2\tilde{\beta}(s,k)},~~~~~\tilde{C}^{\Lambda}_{\pi\pi}(s,k) =  \frac{\tilde{\beta}(s,k)}{2},
\end{equation}
whereas in real space we have
\begin{eqnarray}  
\tilde{C}^{\Lambda}_{\phi\phi}(s,x) &\equiv& \bra{\tilde{\Psi}^{\Lambda}(s)} \phi(x)\phi(0)\ket{\tilde{\Psi}^{\Lambda}(s)} \\
&=& \int_{\mathbb{R}} \frac{dk}{2\pi} ~  e^{i kx}\tilde{C}^{\Lambda}_{\phi\phi}(s,k),~~~\\
\tilde{C}^{\Lambda}_{\pi\pi}(s,x) &\equiv& \bra{\tilde{\Psi}^{\Lambda}(s)} \pi(x)\pi(0)\ket{\tilde{\Psi}^{\Lambda}(s)} \\
&=& \int_{\mathbb{R}} \frac{dk}{2\pi} ~  e^{ikx} \tilde{C}^{\Lambda}_{\pi\pi}(s,k).~~~~~ 
\end{eqnarray}

\vspace{0.5cm}
\begin{center}
    \textit{B. Comparison between pictures}
\end{center}
\vspace{0.5cm}

Consider a one-parameter family of cMERA states
\begin{equation} \label{eq:cMERA_line}
    \ket{\Psi^{\Lambda}(s)} \equiv \exp \left(- i s\left(L+K \right) \right) \ket{\Lambda},
\end{equation}
where $L$ is the non-relativistic rescaling operator 
\begin{eqnarray}
   L &\equiv& \frac{1}{2}\int_{\mathbb{R}} dk [\pi(-k) (k \partial_k + 1/2)\phi(k) + h.c], \\
   && e^{isL}\phi(k)e^{-isL} = e^{s/2}\phi(e^{s}k), \\
&&    e^{isL}\pi(k)e^{-isL} = e^{s/2}\pi(e^{s}k),
\end{eqnarray}
and $K$ is a quasi-local, scale-independent \textit{entangler}
\begin{eqnarray} \label{eq:K}
K &\equiv& \frac{1}{2}\int_{\mathbb{R}_{2}} dxdy ~g(x-y) [\pi(x)\phi(y) + \phi(x)\pi(y)]~~~\\
&=&\frac{1}{2}\int_{\mathbb{R}} dk\, g(k) [\pi(-k)\phi(k) + \phi(-k)\pi(k)],
\end{eqnarray}
which is expressed in terms of a scale-independent entangling profile $g(x)$. Here $g(x)$ is equipped with a constant length scale $1/\Lambda$. (What is important at this stage is not that $K$ is indepedent of $s$, which we have chosen for simplicity, but rather that it entangles at a constant length scale $1/\Lambda$. More generally, we could have chosen $g(s,x)$, leading to a scale dependent entangler $K(s)$ as in Ref. \cite{cMERA}, still with constant constant entangling length $1/\Lambda$.)

In the main text we found it convenient to work instead with a re-scaled picture in which the cMERA state $\ket{\tilde{\Psi}^{\Lambda}(s)}$ is given by
\begin{equation}
   \ket{\tilde{\Psi}^{\Lambda}(s)} \equiv e^{isD} \ket{\Psi^{\Lambda}(s)},
\end{equation}
where $D$ is the relativistic rescaling operator $D$,
\begin{eqnarray}
   D &\equiv& \frac{1}{2}\int_{\mathbb{R}} dk [\pi(-k) (k \partial_k + 1)\phi(k) + h.c], \\
   && e^{isD}\phi(k)e^{-isD} = e^{s}\phi(e^{s}k), \\
&&    e^{isD}\pi(k)e^{-isD} = \pi(e^{s}k).
\end{eqnarray}
That leads to the expression in Eq. \eqref{eq:cMERA_def1} provided that the scale-dependent entangler $\tilde{K}(s)$ reads
\begin{eqnarray}
   \tilde{K}(s)  &\equiv& e^{isD} (L + K - D) e^{-isD} \\
    &=& \frac{1}{2}\int_{\mathbb{R}} dk\, \left(g(k)-\frac{1}{2}\right) \times \nonumber\\
    && ~~~~~~~~~e^{isD} \left[\pi(-k)\phi(k) + \phi(-k)\pi(k)\right] e^{-isD} ~~~~~\\
    &=& \frac{1}{2}\int_{\mathbb{R}} dk\, \left(g(k)-\frac{1}{2}\right) \times \nonumber\\
    && ~~~~~~~~~e^{s}[\pi(-e^sk)\phi(e^sk) + \phi(-e^sk)\pi(e^sk)]  \\
    &=& \frac{1}{2}\int_{\mathbb{R}} dk\,  \left(g(e^{-s}k)-\frac{1}{2}\right) \times \nonumber\\
    && ~~~~~~~~~[\pi(-k)\phi(k) + \phi(-k)\pi(k)],
\end{eqnarray}
so that $\tilde{g}(s,k) = g (e^{-s}k)-1/2$. 

\vspace{0.5cm}
\begin{center}
    \textit{C. Relativistic free boson on the line}
\end{center}
\vspace{0.5cm}

The relativistic free boson has Hamiltonian
\begin{eqnarray}
H^{\QFT} \equiv  \frac{1}{2} \int_{\mathbb{R}} dx[ \pi(x)^2 + \left(\partial_x \phi(x)\right)^2 + m^2 \phi(x)^2 ],
\end{eqnarray}
where $m$ is the mass. We can diagonalize this Hamiltonian by first re-expressing it in terms of Fourier fields $\phi(k), \pi(k)$ and in terms of the annihilation operators $b(k)$, namely
\begin{eqnarray}
H^{\QFT} &=&  \frac{1}{2} \int_{\mathbb{R}} dk[ \pi(-k)\pi(k) + \phi(-k)(k^2 + m^2) \phi(k) ]~~~~~\\
&=& \int_{\mathbb{R}} dk~\sqrt{k^2 + m^2}~b(k)^{\dagger} b(k),
\end{eqnarray}
with
\begin{eqnarray}\label{eq:akm}
b(k) &\equiv& \sqrt{\frac{\beta^{\QFT}(k)}{2}} \phi(k) + \frac{i}{\sqrt{2\beta^{\QFT}(k)}} \pi(k),\\
\beta^{\QFT}(k) &\equiv& \sqrt{k^2 + m^2}
\end{eqnarray} 

Then its ground state $\ket{\Psi^{\QFT}}$ is characterized by being annihilated by $b(k)$,
\begin{equation}
    b(k) \ket{\Psi^{\QFT}} = 0,~~~\forall k \in \mathbb{R}.
\end{equation}
Notice that $\ket{\Psi^{\QFT}}$ comes equipped with a momentum scale $m$. This is reflected in the correlation functions. In momentum space they read 
\begin{eqnarray}
 \bra{\Psi^{\QFT}} \phi(k)\phi(k')\ket{\Psi^{\QFT}} &\equiv& \delta(k+k') ~ C^{\QFT}_{\phi\phi}(k),\\
 \bra{\Psi^{\QFT}} \pi(k)\pi(k')\ket{\Psi^{\QFT}} &\equiv& \delta(k+k') ~ C^{\QFT}_{\pi\pi}(k),
\end{eqnarray}
with
\begin{equation}
    C^{\QFT}_{\phi\phi}(k) = \frac{1}{2\beta^{\QFT}(k)},
~~~~~    C^{\QFT}_{\pi\pi}(k) =  \frac{\beta^{\QFT}(k)}{2}. ~~~~
\end{equation}
In real space we have
\begin{eqnarray}  
C^{\QFT}_{\phi\phi}(x) &\equiv& \bra{\Psi^{\QFT}} \phi(x)\phi(0)\ket{\Psi^{\QFT}} \\ 
&=& \int_{\mathbb{R}} \frac{dk}{2\pi}~e^{i kx} C^{\QFT}_{\phi\phi}(k) ~~~~\\
&=& \frac{1}{2\pi} K_0(mx).~~~~~~~~~~~~~~~~~ \label{eq:lineK0}
\end{eqnarray}
and
\begin{eqnarray}
C^{\QFT}_{\pi\pi}(x) &\equiv& \bra{\Psi^{\QFT}} \pi(x)\pi(0)\ket{\Psi^{\QFT}_m} \\ 
&=& \int_{\mathbb{R}} \frac{dk}{2\pi}~e^{i kx}C^{\QFT}_{\pi\pi}(k) ~~~~~~~~~~~~~~~\\
&=& \frac{1}{2\pi} \frac{m^2}{2}\left(K_0(mx) - K_2(mx) \right), \label{eq:lineK0K2}~~~~~ 
\end{eqnarray}
where $K_n(x)$ is the modified Bessel function of second kinds and order $n$, with short distance expansions
\begin{equation}
 K_0(x) \sim -\log(x),~~~K_2(x) \sim \frac{1}{x^2},
\end{equation}
and long distance expansion 
\begin{equation}
    K_n(x) \sim \sqrt{\frac{\pi}{2}}\frac{e^{-x}}{\sqrt{x}} \left(1 + \frac{4n^2-1}{8x} + O\left(\frac{1}{x^2}\right) \right).
\end{equation}
We therefore find the short-distance expansions
\begin{eqnarray}  \label{eq:mx_approx1short}
C^{\QFT}_{\phi\phi}(x) &\sim& -\log(mx), \\
C^{\QFT}_{\pi\pi}(x) &\sim& -\frac{1}{4\pi}\frac{1}{x^2}, \label{eq:mx_approx2short}
\end{eqnarray}
for $mx \ll 1$, both of which diverge in the limit $mx \rightarrow 0$. For later reference, we add the correlator
\begin{equation}\label{eq:mx_approx3short}
    C^{\QFT}_{\partial_x\phi\partial_x\phi}(x) \sim \frac{1}{x^2},
\end{equation}
obtained by taking the second derivative of $C^{\QFT}_{\phi\phi}(x)$ with respect to $x$.
We also have the long distance expansions
\begin{eqnarray}  \label{eq:mx_approx1}
C^{\QFT}_{\phi\phi}(x) &=& \sqrt{\frac{\pi}{2}}  \frac{e^{-mx}}{\sqrt{mx}} \left(1 + O\left(\frac{1}{mx}\right)\right), \\
C^{\QFT}_{\pi\pi}(x) &=& -\sqrt{\frac{\pi}{2}}  \frac{e^{-mx}}{\sqrt{mx}}\left(1 + O\left(\frac{1}{mx}\right)\right), \label{eq:mx_approx2}
\end{eqnarray}
for $mx \gg 1$, indicating that both correlation functions decay exponentially, with correlation length  $1/m$.

\begin{figure}
	\centering
	\includegraphics[width=1.0\linewidth]{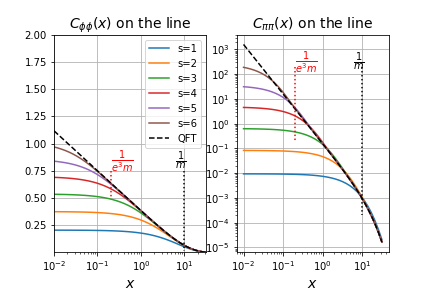}
	\caption{ (Left) Real space correlation functions $C^{\QFT}_{\phi\phi}(x)$ for the ground state of the relativistic free boson on the line with mass $m=0.1$ (discontinuous black line) and $\tilde{C}^{\Lambda}_{\phi\phi}(s,x)$ for the corresponding magic cMERA with $\Lambda=m$ and several values of the scale parameter $s$ (continuous lines of varying color). Notice two characteristic lengths: the IR length $1/m=10$, present both for the ground state of the relativistic free boson and the cMERA, and the UV length $1/(e^s m)$, which is only present in the cMERA and decays exponentially in $s$. The ground state correlator diverges as $x \rightarrow 0$, whereas the cMERA correlators tend to a finite constant for $x$ smaller than the UV length $e^{-s}/m$. (Right) Same for the real space correlation functions $C^{\QFT}_{\pi\pi}(x)$ for the ground state of the relativistic free boson and $\tilde{C}^{\Lambda}_{\pi\pi}(s,x)$ for the magic cMERA.}
	\label{fig:line_corr_2in1}
\end{figure}

\vspace{0.5cm}
\begin{center}
    \textit{D. Magic cMERA on the line}
\end{center}
\vspace{0.5cm}

In the main text we use the magic entangler from Ref. \cite{Yijian}, given by
\begin{equation}
g(k) = \frac{1}{2}\frac{\Lambda^2}{k^2 + \Lambda^2}    
\end{equation}
as an explicit example. It leads to 
\begin{eqnarray}
 \tilde{g}(s,k) = g(e^{-s}k) - \frac{1}{2}= -\frac{1}{2}\frac{(e^{-s}k)^2}{(e^{-s}k)^2 + \Lambda^2}.    
\end{eqnarray}
Then, one can check that
\begin{equation}
    \tilde{\beta}(s,k) = 
    \Lambda  \sqrt{\frac{k^2 + \Lambda^2}{(e^{-s}k)^2 + \Lambda^2}} 
\end{equation}
is a solution of the differential equation \eqref{eq:diff1} with the correct initial conditions $\tilde{\beta}(s=0,k) = \Lambda$.  Notice that $\ket{\tilde{\Psi}^{\Lambda}(s)}$ comes equipped with two momentum scales: $\Lambda$ and $e^{s}\Lambda$, which we can think of as IR and UV scales, respectively. Specifically, its correlation functions in momentum space read 
\begin{eqnarray}
 \bra{\tilde{\Psi}^{\Lambda}(s)} \phi(k)\phi(k')\ket{\tilde{\Psi}^{\Lambda}(s)} &\equiv& \delta(k+k') ~ \tilde{C}^{\Lambda}_{\phi\phi}(s,k),\\
 \bra{\tilde{\Psi}^{\Lambda}(s)} \pi(k)\pi(k')\ket{\tilde{\Psi}^{\Lambda}(s)} &\equiv& \delta(k+k') ~ \tilde{C}^{\Lambda}_{\pi\pi}(s,k),~~~~~
\end{eqnarray}
for
\begin{equation}
    \tilde{C}^{\Lambda}_{\phi\phi}(s,k) = \frac{1}{2\tilde{\beta}(s,k)},~~~~~\tilde{C}^{\Lambda}_{\pi\pi}(s,k) =  \frac{\tilde{\beta}(s,k)}{2}, \label{eq:CCC}
\end{equation}
whereas in real space we have
\begin{eqnarray}  
\tilde{C}^{\Lambda}_{\phi\phi}(s,x) &\equiv& \bra{\tilde{\Psi}^{\Lambda}(s)} \phi(x)\phi(0)\ket{\tilde{\Psi}^{\Lambda}(s)} \\
&=& \int_{\mathbb{R}} \frac{dk}{2\pi} ~  e^{i kx} \tilde{C}^{\Lambda}_{\phi\phi}(s,k),~~~\\
\tilde{C}^{\Lambda}_{\pi\pi}(s,x) &\equiv& \bra{\tilde{\Psi}^{\Lambda}(s)} \pi(x)\pi(0)\ket{\tilde{\Psi}^{\Lambda}(s)} \\ &=& \int_{\mathbb{R}} \frac{dk}{2\pi} ~  e^{i kx} \tilde{C}^{\Lambda}_{\pi\pi}(s,k).~~~~~ 
\end{eqnarray}
Fig. \ref{fig:line_corr_2in1} shows the real-space correlation functions $\tilde{C}^{\Lambda}_{\phi\phi}(s,x)$ and $\tilde{C}^{\Lambda}_{\pi\pi} (s,x)$, obtained through a numerical Fourier transformation, for several values of the scale parameter $s$. One can see that these correlation functions behave differently in three regimes, separated by the UV length $e^{-s}/\Lambda$ and the IR length $1/\Lambda$: (i) for $x \in [0,e^{-s}/\Lambda]$, the correlator is approximately constant, with a value that grows with $s$; (ii) for $x \in [e^{-s}/\Lambda, 1/\Lambda]$, $\tilde{C}^{\Lambda}_{\phi\phi}(s,x)$ and $\tilde{C}^{\Lambda}_{\pi\pi}(s,x)$ scale approximately as $-\log(\mu x)$ and $1/x^2$, respectively; (iii) for $x > 1/\Lambda$, the correlation functions decay exponentially, with correlation length $1/\Lambda$.

\begin{figure}
	\centering
	\includegraphics[width=1.0\linewidth]{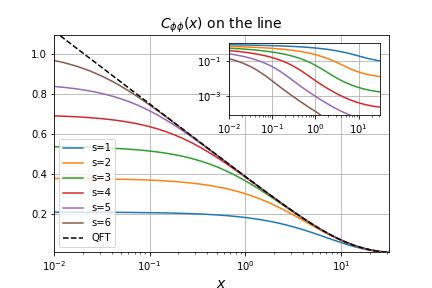}
	\caption{Same as Fig. \eqref{fig:line_corr_2in1}(Left). The inset shows the relative error $E_{\phi\phi}(s,x)$ between the ground state and cMERA correlators, which starts to decay sharply with x for value of $x$ larger than the UV length $x_{\UV} e^{-s}/m $.}
	\label{fig:line_corr}
\end{figure}

\begin{figure}
	\centering
	\includegraphics[width=1.0\linewidth]{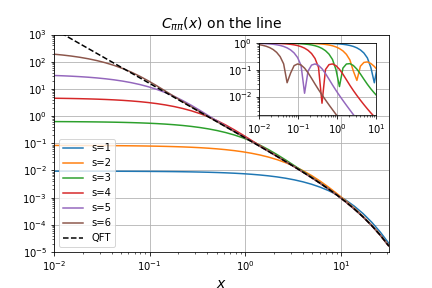}
	\caption{Same as Fig. \eqref{fig:line_corr_2in1}(Right). The inset shows the relative error $E_{\pi\pi}(s,x)$ between the ground state and cMERA correlators, which starts to decay sharply with x for value of $x$ larger than the UV length $x_{\UV} e^{-s}/m $.}
	\label{fig:line_corr_pi}
\end{figure}

We remark that, had we not moved to the re-scaled picture with $\ket{\tilde{\Psi}^{\Lambda}(s)} = e^{isD} \ket{\Psi^{\Lambda}(s)}$ but used $\ket{\Psi^{\Lambda}(s)}$ instead, the corresponding correlation functions $C^{\Lambda}_{\phi\phi}(s,x)$ and $C^{\Lambda}_{\pi\pi}(s,x)$ would have had their UV and IR length scales respectively be $1/\Lambda$ (instead of $e^{-s/\Lambda}$) and $e^{s}/\Lambda$ (instead of $1/\Lambda$).

\vspace{0.5cm}
\begin{center}
    \textit{D. Ground state vs magic cMERA on the line}
\end{center}
\vspace{0.5cm}

We are now ready to compare the ground state $\ket{\Psi^{QFT}}$ of the relativistic free boson with the magic cMERA $\ket{\Psi^{\Lambda}(s)}$ of Ref. \cite{Yijian}. They are annihilated, respectively, by operators $b(k)$ and $\tilde{b}^{\Lambda}(s,k)$, given by functions $\beta^{\QFT}(k)$ and $\tilde{\beta}(s,k)$:
\begin{eqnarray}
 \beta^{\QFT}(k) &=& \sqrt{k^2 + m^2} ~~~~~~~~~~\\ 
 \tilde{\beta}(s,k) &=&\sqrt{k^2 + \Lambda^2} \frac{1}{\sqrt{(e^{-s}k/\Lambda)^2 + 1}} ~~~ \nonumber~~\\
  &=& \sqrt{k^2 + \Lambda^2} \left(1 + O\left(\left(\frac{e^{-s}k}{m}\right)^2\right)\right).
\end{eqnarray}  
We thus see that if we choose $\Lambda=m>0$ (in the massless case $m=0$, this requires introducing a small mass $m>0$ as an IR regulator), then we find that
\begin{eqnarray}
\tilde{\beta}(s,k)
 &=& \sqrt{k^2 + m^2} \frac{1}{\sqrt{(e^{-s}k/m)^2 + 1}} \\
 &=& \beta^{\QFT}(k) \left(1 + O\left(\left(\frac{e^{-s}k}{m}\right)^2\right) \right). \label{eq:betabeta1}
\end{eqnarray}
That is, the cMERA annihilation operators are similar to the ground state annihilation operators for momenta $k$ smaller than the UV momentum $e^{s}m$. More specifically, Eqs. \eqref{eq:CCC} and
\eqref{eq:betabeta1} together imply 
\begin{eqnarray}
 \tilde{C}^{\Lambda}_{\phi\phi}(s,k) &=& C^{\QFT}_{\phi\phi}(k) \left(1 + O\left( \left(\frac{e^{-s}k}{m}\right)^2\right) \right),~~ \\
 \tilde{C}^{\Lambda}_{\pi\pi}(s,k) &=& C^{\QFT}_{\pi\pi}(k) \left(1 + O\left( \left(\frac{e^{-s}k}{m}\right)^2\right) \right),  
\end{eqnarray}
indicating that the cMERA correlators match those of the QFT ground state at small momenta. Then, through a Fourier transform to real space we find, 
\begin{eqnarray} \label{eq:CLamQFT1}
 \tilde{C}^{\Lambda}_{\phi\phi}(s,x) &=& C^{\QFT}_{\phi\phi}(x) + O\left(\frac{e^{-2s}}{m^2} \frac{d^2}{dx^2} C^{\QFT}_{\phi\phi}(x) \right),   \\
 \tilde{C}^{\Lambda}_{\pi\pi}(s,k) &=&  C^{\QFT}_{\pi\pi}(x) + O\left(\frac{e^{-2s}}{m^2} \frac{d^2}{dx^2} C^{\QFT}_{\pi\pi}(x) \right).\label{eq:CLamQFT2} ~~~~
\end{eqnarray}

Next we consider two regimes for which we can derive approximate expressions.

Firstly, for $m x\ll 1$, but assuming $x \gg e^{-s}/m$ (that is, $k \ll e^{s}m$), from the short distance expansions \eqref{eq:mx_approx1short}-\eqref{eq:mx_approx2short}, we obtain
\begin{eqnarray}  \label{eq:mx_dd1short}
\frac{d^2}{dx^2}C^{\QFT}_{\phi\phi}(x) \sim \frac{1}{x^2},~~~
\frac{d^2}{dx^2}C^{\QFT}_{\pi\pi}(x) \sim \frac{1}{x^4}, \label{eq:mx_dd2short}
\end{eqnarray}
and therefore
\begin{eqnarray} \label{eq:approx_critical1}
 \tilde{C}^{\Lambda}_{\phi\phi}(s,x) &=& C^{\QFT}_{\phi\phi}(x)\left(1 + O\left(\frac{e^{-2s}}{(mx)^2\log(mx)} \right)\right),  ~~~~~~~ \\
 \tilde{C}^{\Lambda}_{\pi\pi}(s,k) &=&  C^{\QFT}_{\pi\pi}(x)\left(1 + O\left(\frac{e^{-2s}}{(mx)^2} \right)\right). ~~~~\label{eq:approx_critical2}
\end{eqnarray}
Similarly, we would find
\begin{equation}
    \tilde{C}^{\Lambda}_{\partial_x \phi \partial_x\phi}(s,k) =  C^{\QFT}_{\partial_x \phi \partial_x\phi}(x)\left(1 + O\left(\frac{e^{-2s}}{(mx)^2} \right)\right). \label{eq:approx_critical3}
\end{equation}
Recall that this is the setting relevant to the massless case, $m=0$, where we introduce a small mass $m$ as an IR regulator. Thus, the cMERA succeeds at recovering the power-law decay (or, for $C_{\phi\phi}^{\QFT}(x)$, the logarithmic scaling) characteristic of the correlation functions of the free boson conformal field theory, provided that $x \in [e^{-s}/m, 1/m]$ is not too close to the UV length $e^{-s}/m$ and IR length $1/m$.

Secondly, for $x \gg 1/m$,  Eqs. \eqref{eq:mx_approx1}-\eqref{eq:mx_approx2} imply
\begin{eqnarray}  \label{eq:mx_dd1}
\frac{d^2}{dx^2}C^{\QFT}_{\phi\phi}(x) &=& m^2~ C^{\QFT}_{\phi\phi}(x) \left(1 + O\left(\frac{1}{mx}\right)\right),~~~\\
\frac{d^2}{dx^2}C^{\QFT}_{\pi\pi}(x) &=& m^2~ C^{\QFT}_{\pi\pi}(x) \left(1 + O\left(\frac{1}{mx}\right)\right), \label{eq:mx_dd2}
\end{eqnarray}
and therefore
\begin{eqnarray}\label{eq:approx_massive1}
 \tilde{C}^{\Lambda}_{\phi\phi}(s,x) &=& C^{\QFT}_{\phi\phi}(x)\left(1 + O\left(e^{-2s} \right)\right),   \\
 \tilde{C}^{\Lambda}_{\pi\pi}(s,x) &=&  C^{\QFT}_{\pi\pi}(x)\left(1 + O\left(e^{-2s} \right)\right). ~~~~\label{eq:approx_massive2}
\end{eqnarray}

Expressions \eqref{eq:approx_critical1}-\eqref{eq:approx_critical3} and \eqref{eq:approx_massive1}-\eqref{eq:approx_massive2} will be useful in the next appendix, when trying to characterize the accuracy of cMERA for the ground state of the relativistic boson on the circle.

Notice that \eqref{eq:approx_critical2}-\eqref{eq:approx_critical3} and \eqref{eq:approx_massive1}-\eqref{eq:approx_massive2} imply that, for $x$ in the relevant regime of values, the relative error between cMERA and ground state correlators on the line is exponentially suppressed with $s$, 
\begin{eqnarray} \label{eq:E_rel_phi}
 E_{\phi\phi}(x) &\equiv& \left|\frac{\tilde{C}^{\Lambda}_{\phi\phi}(s,k) - C^{\QFT}_{\phi\phi}(x) }{C^{\QFT}_{\phi\phi}(x) }\right| = O\left( e^{-2s}\right),~~~~~\\
 E_{\pi\pi}(x) &\equiv& \left|\frac{\tilde{C}^{\Lambda}_{\pi\pi}(s,k) - C^{\QFT}_{\pi\pi}(x) }{C^{\QFT}_{\pi\pi}(x) }\right| = O\left( e^{-2s}\right). \label{eq:E_rel_pi}
\end{eqnarray}

Finally, the insets in Figs. \eqref{fig:line_corr} and \eqref{fig:line_corr_pi} show the actual errors $E_{\phi\phi}(s,x)$ and $E_{\pi\pi}(s,x)$ computed numerically.


\vspace{0.5cm}
\textbf{Appendix A2: Entangling evolution on the circle}
\vspace{0.5cm}

In this appendix we provide a more detailed derivation of several of the expressions for the cMERA on the circle that were used in the main text. 

\vspace{0.5cm}
\begin{center}
    \textit{A. cMERA on the circle}
\end{center}
\vspace{0.5cm}

We consider a bosonic field on a circle of size $l_c$, with field operators $\phi(x),\pi(x)$ for $x\in[0,l_c)$ and canonical commutation relations $[\phi(x),\pi(y)]=i\delta(x-y)$. We introduce a discrete set of Fourier modes
\begin{eqnarray}
\phi(n) &\equiv& \frac{1}{\sqrt{l_c}} \int_0^{l_c} dx ~e^{-i k_{n}x} \phi(x), \\
\pi(n) &\equiv&  \frac{1}{\sqrt{l_c}} \int_0^{l_c} dx ~e^{-i k_{n}x}  \pi(x),
\end{eqnarray}
with $n \in \mathbb{Z}$ and $k_n \equiv 2\pi n/l_c$. They obey $\phi(n)^{\dagger} = \phi(-n)$, $\pi(n)^{\dagger} = \pi(-n)$ and commutation relations $[\phi(n), \pi(n)]= i \delta_{n,-m}$. In terms of  $\phi(n), \pi(n)$, the original fields $\phi(x),\pi(x)$ read
\begin{eqnarray}
\phi(x) = \frac{1}{\sqrt{l_c}} \sum_n e^{i k_{n}x} \phi(n),\\
\pi(x) = \frac{1}{\sqrt{l_c}} \sum_n e^{i k_{n}x} \pi(n),
\end{eqnarray}
where we have used 
\begin{eqnarray}
    \int_0^{l_c} e^{i(k_{n}-k_{m})x} dx&=& l_c \delta_{n,m},   \\  
    \sum_{n\in \mathbb{Z}} e^{ik_{n}(x-y)}&=&l_c\delta(x-y),
\end{eqnarray}
where the last expression follows from $\sum_{n} e^{iaxn} = \frac{2\pi}{a}\delta(x)$.

Consider an entangler $\tilde{K}_c(s)$ of the form
\begin{eqnarray}
 \tilde{K}_c(s) &\equiv& \frac{1}{2}\!\int_{0}^{l_c}\!\!\! dxdy ~\tilde{g}_c(s,x\!-\!y)[\pi(x)\phi(y)\! + \!\phi(x)\pi(y)]~~~~~~~~\\
&=& \frac{1}{2} \sum_{n\in \mathbb{N}} \tilde{g}_c(s,n) [\pi(-n)\phi(n) + \phi(-n)\pi(n)],
\end{eqnarray}
where  $\tilde{g}_c(s,x)$ is an entangling profile and $\tilde{g}_c(s,n) \equiv \int_{0}^{l_c} dx~e^{-ik_{n}x} \tilde{g}_c(s,x)$ is its discrete Fourier transform. We define the cMERA state $\ket{\tilde{\Psi}_c^{\Lambda}(s)}$ as an entangling evolution by $\tilde{K}_c(s)$,
\begin{equation} \label{eq:cMERAtilde_circle1}
    \ket{\tilde{\Psi}_c^{\Lambda}(s)} \equiv \mathcal{P}\exp{\left(-i \int_0^{s} ds'~\tilde{K}_c(s')\right)} \ket{\Lambda_c},
\end{equation}
starting from the unentangled state $\ket{\Lambda_c}$ given by
\begin{eqnarray}\label{eq:Lambda_c1}
&& \tilde{b}^{\Lambda}(0,n) \ket{\Lambda_c} = 0,~~~~ \forall n \in \mathbb{Z}, \\
&& \tilde{b}^{\Lambda}(0,n) \equiv \sqrt{\frac{\Lambda}{2}} \phi(n) + \frac{i}{\sqrt{2\Lambda}} \pi(n). \label{eq:initial_R2}
\end{eqnarray} 

In order to characterize the cMERA state $\ket{\tilde{\Psi}_c^{\Lambda}(s)}$ in terms of a complete set of annihilation operators $\tilde{b}^{\Lambda}(s,n)$,
\begin{equation}
    \tilde{b}^{\Lambda}(s,n) \ket{\Psi_c^{\Lambda}(s)} = 0, ~~~\forall n \in \mathbb{Z},
\end{equation}
we assume an expression of the form
\begin{equation} \label{eq:ansatz_bn}
    \tilde{b}^{\Lambda}(s,n) \equiv \sqrt{\frac{\tilde{\beta}_c(s,n)}{2}}\phi(n) + \frac{i}{\sqrt{2\tilde{\beta}_c(s,n)}}\pi(n),
\end{equation}
and require that the operators $\tilde{b}^{\Lambda}(s,n)$ change with $s$ according to the entangling evolution, that is
\begin{equation} \label{eq:evolve_scale2}
    \frac{\partial}{\partial s} \tilde{b}^{\Lambda}(s,n) = -i [\tilde{K}_c(s), \tilde{b}^{\Lambda}(s,n)],~~~~\forall n \in \mathbb{Z},
\end{equation}
with Eq. \eqref{eq:initial_R2} as the initial conditions at $s=0$. Notice that Eq. \eqref{eq:evolve_scale2} is formally equivalent to Eq.
\eqref{eq:evolve_scale1}. Following the derivation after Eq.
\eqref{eq:evolve_scale1} we conclude that $\tilde{\beta}_c(s,n)$ must obey the differential equation
\begin{equation} \label{eq:diff2}
    \frac{\partial}{\partial s} \tilde{\beta}_c(s,n) = -2\tilde{g}_c(s,n) \tilde{\beta}_c(s,n),~~~~\forall n \in \mathbb{Z},
\end{equation}
with initial condition (cf. Eq. \eqref{eq:initial_R2})
\begin{equation} 
    \tilde{\beta}_c(s=0,n) = \Lambda,~~~~\forall n \in \mathbb{Z}.
\end{equation}

\vspace{0.5cm}
\begin{center}
    \textit{B. Relativistic free boson on the circle}
\end{center}
\vspace{0.5cm}

The relativistic free boson on a circle has Hamiltonian
\begin{eqnarray}
H^{\QFT}_{c} \equiv  \frac{1}{2} \int_{0}^{l_c} dx[ \pi(x)^2 + \left(\partial_x \phi(x)\right)^2 + m^2 \phi(x)^2 ],~~~
\end{eqnarray}
where $m$ is the mass and $l_c$ is the circle's perimeter. As on the line, we diagonalize $H^{\QFT}_{c}$ by first re-expressing it in terms of Fourier fields $\phi(n), \pi(n)$ and then in terms of the annihilation operators $b_m(n)$, namely
\begin{eqnarray}
H^{\QFT}_{c} &=&  \frac{1}{2} \sum_{n \in \mathbb{Z}} [ \pi(-n)\pi(n) + \phi(-n)(k_n^2 + m^2) \phi(n) ]~~~~~~~\\
&=& \sum_{n\in \mathbb{Z}}~\sqrt{k_n^2 + m^2}~b(n)^{\dagger} b(n),
\end{eqnarray}
with $k_n \equiv 2\pi n/l_c$ and
\begin{eqnarray}\label{eq:akm}
b(n) &\equiv& \sqrt{\frac{\beta^{\QFT}_{c}(n)}{2}} \phi(n) + \frac{i}{\sqrt{2\beta^{\QFT}_{c}(n)}} \pi(n),~~~~~~~\\
\beta^{\QFT}_{c}(n) &\equiv& \sqrt{k_n^2 + m^2}
\end{eqnarray} 

\begin{figure}
	\centering
	\includegraphics[width=1.0\linewidth]{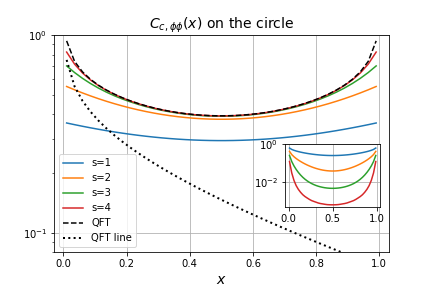}
	\caption{Real space correlation functions $C^{\QFT}_{c,\phi\phi}(x)$ for the ground state of the relativistic free boson with mass $m=1$ on a circle of length $l_c=1$ (discontinuous black line) and $\tilde{C}^{\Lambda}_{c,\phi\phi}(s,x)$ for the corresponding magic cMERA with $\Lambda=m$ and several values of the scale parameter $s$ (continuous lines of varying color). The inset shows the relative error $E_{c,\phi\phi}(s,x)$ between the ground state and cMERA correlators, which starts to decay sharply with x for value of $x$ larger than the UV length $x_{\UV} e^{-s}/m $.}
	\label{fig:circle_corr}
\end{figure}

Then its ground state $\ket{\Psi^{\QFT}_{c}}$ is characterized by being annihilated by $b(n)$,
\begin{equation}
    b(n) \ket{\Psi^{\QFT}_{c}} = 0,~~~\forall n \in \mathbb{Z}.
\end{equation}
The correlators in momentum space read 
\begin{eqnarray}
 \bra{\Psi_{c}^{\QFT}} \phi(n)\phi(n')\ket{\Psi_{c}^{\QFT}} &\equiv& \delta_{n,-n'} ~ C^{\QFT}_{c,\phi\phi}(n),~~~~~\\
 \bra{\Psi_{c}^{\QFT}} \pi(n)\pi(n')\ket{\Psi_{c}^{\QFT}} &\equiv& \delta_{n,-n'}~ C^{\QFT}_{c,\pi\pi}(n),
\end{eqnarray}
with
\begin{equation}
    C^{\QFT}_{c,\phi\phi}(n) = \frac{1}{2\beta_{c}^{\QFT}(n)},
~~~~~    C^{\QFT}_{c,\pi\pi}(n) =  \frac{\beta^{\QFT}_{c}(n)}{2}. ~~~~
\end{equation}
and, in real space,
\begin{eqnarray}  
C^{\QFT}_{c,\phi\phi}( x) &\equiv& \bra{\Psi^{\QFT}_{c}} \phi(x)\phi(0)\ket{\Psi^{\QFT}_c} \\
&=&  \frac{1}{l_c}\sum_{n\in \mathbb{Z}} e^{ik_nx} C_{c,\phi\phi}^{\QFT}(n), ~~~~\\
C^{\QFT}_{c,\pi\pi}(x) &\equiv& \bra{\Psi_c^{\QFT}} \phi(x)\phi(0)\ket{\Psi_c^{\QFT}} \\
&=& \frac{1}{l_c} \sum_{n \in \mathbb{Z}} e^{i k_nx} C_{c,\pi\pi}^{\QFT}(n).~~~~
\end{eqnarray}

\begin{figure}
	\centering
	\includegraphics[width=1.0\linewidth]{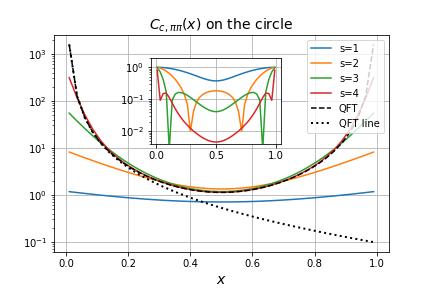}
	\caption{Real space correlation functions $C^{\QFT}_{c,\pi\pi}(x)$ for the ground state of the relativistic free boson with mass $m=1$ on a circle of length $l_c=1$ (discontinuous black line) and $\tilde{C}^{\Lambda}_{c,\pi\pi}(s,x)$ for the corresponding magic cMERA with $\Lambda=m$ and several values of the scale parameter $s$ (continuous lines of varying color). The inset shows the relative error $E_{c,\pi\pi}(s,x)$ between the ground state and cMERA correlators, which starts to decay sharply with x for value of $x$ larger than the UV length $x_{\UV} e^{-s}/m $.}
	\label{fig:circle_corr_pi}
\end{figure}

Using the method of images (reviewed in Appendix [A3]) we can then relate the real space correlators on the circle to those on the line through
\begin{eqnarray}  \label{eq:circle_line1}
C^{\QFT}_{c,\phi\phi}(x) = \sum_{n\in \mathbb{Z}} C^{\QFT}_{\phi\phi}(x+n l_c), \\
C^{\QFT}_{c,\pi\pi}(x) = \sum_{n\in \mathbb{Z}}C^{\QFT}_{\pi\pi}(x + n l_c), \label{eq:circle_line2}
\end{eqnarray}
and thus express them in terms of the explicit solutions (modified Bessel functions of the second kind $K_0(mx)$ and $K_2(mx)$, see Eqs. \eqref{eq:lineK0} and \eqref{eq:lineK0K2}) and corresponding short distance and long distance expansions we found on the line.

In particular, if we consider the short distance expansion (valid for $mx \ll 1$), and artificially expand it to arbitrary values of $x \in \mathbb{R}_{+}$, we obtain
\begin{eqnarray}  \label{eq:circle_CFT}
C^{\QFT}_{c,\phi\phi}(x) &\sim& - \sum_{n\in \mathbb{Z}} \log(m|x+n l_c|),~~~~~ \\
C^{\QFT}_{c,\pi\pi}(x) &\sim& - \sum_{n\in \mathbb{Z}} \frac{1}{(x+n l_c)^2}. \label{eq:circle_CFT2}
\end{eqnarray}
The sum for $C^{\QFT}_{c,\phi\phi}(x)$ diverges for all $x>0$, but the sum for $C^{\QFT}_{c,\pi\pi}(x)$ is finite, as is also the (equivalent) sum for $C^{\QFT}_{c,\partial_x \phi \partial_x \phi}(x)$, see. Eq. \eqref{eq:mx_approx3short}. This is consistent with the well-known fact that, in the free boson CFT, the field $\phi(x)$ is not necessarily a well-defined object, but the fields $\pi(x)$ and $\partial_x\phi(x)$ are.

On the other hand, if we assume that the circle is large compared to the correlation length, that is $m l_c \gg 1$, and consider a value of $x \in [0, l_c]$ such that the long distance expansion is valid (that is, $m|x+nl_c| \gg 1$), we obtain for both $C^{\QFT}_{c,\phi\phi}(x)$ and $C^{\QFT}_{c,\pi\pi}(x)$ a sum of the form
\begin{equation} 
\sqrt{\frac{\pi}{2}}  \sum_{n\in \mathbb{Z}} \frac{e^{-m|x+nl_c|}}{\sqrt{m|x+nl_c|}}\left(1 + O\left(\frac{1}{m|x+nl_c|}\right) \right), ~~~
\end{equation}
where the terms in the sum are exponentially suppressed with growing values of $|n|$ and therefore converges.

\vspace{0.5cm}
\begin{center}
    \textit{C. Example: magic cMERA on the circle}
\end{center}
\vspace{0.5cm}

In the main text we have adjusted the magic entangler of Ref. \cite{Yijian} (originally proposed for the line) to the circle, with real space entangling profile
\begin{eqnarray}
g_c(s,x)  &\equiv& \frac{e^{s}\Lambda}{4}\sum_{n\in \mathbb{Z}} e^{-e^{s}\Lambda|x+n l_c|}~~~~~\\
 &=& \frac{e^{s}\Lambda}{4} \frac{\cosh\left(e^{s}\Lambda\left(\frac{l_c}{2}-x\right)\right)}{\sinh\left(e^{s}\Lambda \frac{l_c}{2}\right)}
\end{eqnarray}
where we used that, for $a\in [0,b]$,
\begin{eqnarray}
 \sum_{n\in \mathbb{Z}} e^{-|a+n b|} &=& \left(e^{-a}+e^{-(b-a)}\right) \sum_{n=0}^{\infty} e^{-nb}\\
 &=& \left(e^{\frac{b}{2}-a} + e^{-\left(\frac{b}{2}-a\right)} \right) \frac{e^{-\frac{b}{2}}}{1-e^{-b}}~~~~~\\
 &=& \frac{\cosh\left(\frac{b}{2}-a\right)}{\sinh\left(\frac{b}{2}\right)},
\end{eqnarray}
and momentum space profile
\begin{equation}
    \tilde{g}_c(s,n) = -\frac{1}{2} \frac{(e^{-s} k_n)^2}{(e^{-s} k_n)^2 + \Lambda^2}. 
\end{equation}
This results in the solution $\tilde{\beta}_c(s,n)$ of the differential equation \eqref{eq:diff2} given by
\begin{eqnarray}
 \tilde{\beta}_c(s,n) = \Lambda \sqrt{\frac{k_n^2 + \Lambda^2}{(e^{-s}k_n)^2 + \Lambda^2}}, 
\end{eqnarray}
which indeed fulfils the initial condition \eqref{eq:initial_R2}. The resulting cMERA has correlators 
\begin{eqnarray}
 \bra{\tilde{\Psi}^{\Lambda}_c(s)} \phi(n)\phi(n')\ket{\tilde{\Psi}^{\Lambda}_c(s)} &=& \delta_{n,-n'} \tilde{C}^{\Lambda}_{c,\phi\phi}(s,n),\\
 \bra{\tilde{\Psi}^{\Lambda}_c(s)} \pi(n)\pi(n')\ket{\tilde{\Psi}^{\Lambda}_c(s)} &=& \delta_{n,-n'} \tilde{C}^{\Lambda}_{c,\pi\pi}(s,n),~~~~~
\end{eqnarray}
for
\begin{equation}
    \tilde{C}^{\Lambda}_{c,\phi\phi}(s,n) = \frac{1}{2\tilde{\beta}_c(s,n)},~~~\tilde{C}^{\Lambda}_{c,\pi\pi}(s,n) =  \frac{\tilde{\beta}_c(s,n)}{2},~~~
\end{equation}
or, in real space,
\begin{eqnarray}  
\tilde{C}^{\Lambda}_{c,\phi\phi}(s,x) &\equiv& \bra{\tilde{\Psi}^{\Lambda}_c(s)} \phi(x)\phi(0)\ket{\tilde{\Psi}^{\Lambda}_c(s)} \\
&=&  \frac{1}{l_c}\sum_{n\in \mathbb{Z}} e^{ik_nx} \tilde{C}^{\Lambda}_{c,\phi\phi}(n), ~~~~\\
\tilde{C}^{\Lambda}_{c,\pi\pi}(s,x) &\equiv& \bra{\tilde{\Psi}^{\Lambda}_c(s)} \phi(x)\phi(0)\ket{\tilde{\Psi}^{\Lambda}_c(s)} \\
&=& \frac{1}{l_c} \sum_{n \in \mathbb{Z}} e^{i k_nx} \tilde{C}^{\Lambda}_{c,\pi\pi}(n).~~~~
\end{eqnarray}

Using the method of images (reviewed in Appendix [A3]) we can also relate the real space correlators on the circle to those on the line through
\begin{eqnarray}  \label{eq:circle_line1_Lam}
\tilde{C}^{\Lambda}_{c,\phi\phi}(s,x) &=& \sum_{n\in \mathbb{Z}} \tilde{C}^{\Lambda}_{\phi\phi}(s, x+n l_c), \\
\tilde{C}^{\Lambda}_{c,\pi\pi}(s,x) &=& \sum_{n\in \mathbb{Z}} \tilde{C}^{\Lambda}_{\pi\pi}(s,x + n l_c), \label{eq:circle_line2_Lam}
\end{eqnarray}

\vspace{0.5cm}
\begin{center}
    \textit{C. Ground state vs magic cMERA on the circle}
\end{center}
\vspace{0.5cm}

The comparison between the ground state $\ket{\Psi_{c}^{\QFT}}$ of the relativistic free boson and the magic cMERA $\ket{\tilde{\Psi}_c^{\Lambda}(s)}$ on the circle can be conducted, thanks to the method of images, by simply translating expressions from the line to the circle. In particular, combining Eqs. \eqref{eq:CLamQFT1}-\eqref{eq:CLamQFT2} with \eqref{eq:circle_line1_Lam}-\eqref{eq:circle_line2_Lam} we readily obtain
\begin{eqnarray} \label{eq:CLamQFT1c}
 \tilde{C}^{\Lambda}_{c,\phi\phi}(s,x) &=& C^{\QFT}_{c,\phi\phi}(x) + O\left(\frac{e^{-2s}}{m^2} \frac{d^2}{dx^2} C^{\QFT}_{c,\phi\phi}(x) \right),   \\
 \tilde{C}^{\Lambda}_{c,\pi\pi}(s,k) &=&  C^{\QFT}_{c,\pi\pi}(x) + O\left(\frac{e^{-2s}}{m^2} \frac{d^2}{dx^2} C^{\QFT}_{c,\pi\pi}(x) \right).\label{eq:CLamQFT2c} ~~~~~~
\end{eqnarray}

Then, using the $mx \ll 1$ expansion (for the massless case, with $m>0$ an IR regulator to be sent to zero) we find that Eqs. \eqref{eq:approx_critical2}-\eqref{eq:approx_critical3}
imply that
\begin{eqnarray}
 \tilde{C}^{\Lambda}_{c,\pi\pi}(s,k) &=&  C^{\QFT}_{c,\pi\pi}(x) \left(1 + O\left( e^{-2s}\right) \right), \\
  \tilde{C}^{\Lambda}_{c,\partial_x \phi \partial_x \phi}(s,k) &=&  C^{\QFT}_{c,\partial_x \phi \partial_x \phi}(x) \left(1 + O\left( e^{-2s}\right) \right).
\end{eqnarray}
Similarly, for $m l_c \gg 1$ (circle of large length compared to correlation length), for $x \in [0, l_c]$ such that the long distance expansion is valid (that is, $m|x+nl_c| \gg 1$), Eqs. \eqref{eq:approx_massive1}-\eqref{eq:approx_massive2} imply
\begin{eqnarray}
 \tilde{C}^{\Lambda}_{c,\phi\phi}(s,k) &=&  C^{\QFT}_{c,\phi\phi}(x) \left(1 + O\left( e^{-2s}\right) \right), \\
  \tilde{C}^{\Lambda}_{c,\pi\pi}(s,k) &=&  C^{\QFT}_{c,\pi\pi}(x) \left(1 + O\left( e^{-2s}\right) \right).
\end{eqnarray}

The above expressions imply that, for $x$ in the relevant regime of values, the relative error between cMERA and ground state correlators on the circle is exponentially suppressed with $s$, e.g.

\begin{eqnarray} \label{eq:E_rel_phic}
 E_{c,\phi\phi}(s,x) &\equiv& \left|\frac{\tilde{C}^{\Lambda}_{c,\phi\phi}(s,x) - C^{\QFT}_{c,\phi\phi}(x) }{C^{\QFT}_{c,\phi\phi}(x) }\right| = O\left( e^{-2s}\right),~~~~~\\
 E_{c,\pi\pi}(s,x) &\equiv& 
 \left|\frac{\tilde{C}^{\Lambda}_{c,\pi\pi}(s,x) - C^{\QFT}_{c,\pi\pi}(x) }{C^{\QFT}_{c,\pi\pi}(x) }\right|  = O\left( e^{-2s}\right). \label{eq:E_rel_pic}
\end{eqnarray}

More generally, consider a cMERA correlator $C^{\Lambda}(x)$ that approximates a QFT ground state correlator $C^{\QFT}(x)$ on the line at distances larger than $x_{UV}$, in the sense that there is a small $\epsilon>0$ such that for any $|x| \geq x_{UV}$,
\begin{equation}
\left|\frac{C^{\Lambda}(x) - C^{\QFT}(x)}{C^{\QFT}(x)}\right| \leq \epsilon.
\end{equation}
We will also assume that the QFT correlator does not change sign for $x\geq x_{UV}$, say $C^{\QFT}(x) \geq 0 $ for all $x \geq x_{UV}$. Consider now a cMERA and ground state on a circle of size $l_c$ obtained by the method of images (see Appendix [A3]), so that their correlators $C^{\Lambda}_{c}(x)$ and $C^{\QFT}_c(x)$ read
\begin{eqnarray}
 C^{\Lambda}_{c}(x) &=& \sum_{n\in \mathbb{Z}} C^{\Lambda}(x+n l_c), \nonumber\\ 
C^{\QFT}_c(x) &=& \sum_{n\in \mathbb{Z}} C^{\QFT}(x+n l_c).  
\end{eqnarray}
Then, for any $x\in [x_{UV}, l_c/2]$ (that is, such that $|x+nl_c| \geq x_{UV}$) we have that also
 \begin{equation}
\left|\frac{C_c^{\Lambda}(x) - C_c^{\QFT}(x)}{C_c^{\QFT}(x)}\right| \leq \epsilon,
\end{equation}
so that also on the circle the cMERA correlator approximates the ground state correlator.

Indeed, we have that
\begin{eqnarray}
 &&\left|C_c^{\Lambda}(x) - C_c^{\QFT}(x)\right| \\
 &=&  \left|\sum_{n\in \mathbb{Z}} C^{\Lambda}(x+nl_c) - C^{\QFT}(x+nl_c)\right|~~~~ \\
 &\leq& \sum_{n\in \mathbb{Z}} \left|C^{\Lambda}(x+nl_c) - C^{\QFT}(x+nl_c)\right| \\
 &\leq& \sum_{n\in \mathbb{Z}} \epsilon \left|C^{\QFT}(x+nl_c)\right|\\
 &=& \epsilon  \left|\sum_{n\in \mathbb{Z}} C^{\QFT}(x+nl_c)\right|\\
 &=& \epsilon \left| C_c^{\QFT}(x)\right|,
\end{eqnarray}
where in the last step we used that, by assumption, $C^{\QFT}(x+nl_c)$ does not change sign as a function of $n$.

\vspace{0.5cm}
\textbf{Appendix A3: The method of images}
\vspace{0.5cm}

In this Appendix we review the method of images, specialized to connecting the Fourier transforms of two closely related functions $f(x)$ and $f_c(x)$, one the line and the other on the circle. First we state the main result and provide a proof. Then we review in detail how this applies to several examples of interest in the main text.

\vspace{0.5cm}
\begin{center}
    \textit{A. Relating the line and the circle}
\end{center}
\vspace{0.5cm}

Let us consider a function $f(x)$ defined on the real line,  with $f(k)$ its Fourier transform,
\begin{eqnarray}
    f(k) &\equiv& \int_{\mathbb{R}} dx~e^{-ikx}f(x),~~~\\
    f(x) &=& \frac{1}{2\pi}\int_{\mathbb{R}}dk ~e^{ikx} f(k).
\end{eqnarray}
Let us also consider a function $f_{c}(x)$ on a circle of size $l_c$, with $f_c(n)$ its discrete Fourier transform,
\begin{eqnarray}
    f_c(n) &\equiv& \int_{0}^{l_c} dx~e^{-ik_nx}f_{c}(x),~~~\\
    f_c(x) &=& \frac{1}{l_c} \sum_{n\in \mathbb{Z}} e^{ik_nx} f_c(n),
\end{eqnarray}
where $k_n \equiv 2\pi n/l_c$. 

We will assume that $f(x)$ decays sufficiently fast with $|x|$ for large $|x|$, so that for any $x \in [0,l_c)$, the sum $\sum_{n\in \mathbb{Z}} f(x+nl_c)$ is finite.

\vspace{0.5cm}

\textbf{Theorem (method of images)}: 
\begin{eqnarray}
f_{c}(x) &=& \sum_{n\in \mathbb{Z}} f(x+nl_c),~~~~\forall x\in [0,l_c) \\
 \Longleftrightarrow ~~~~~   f_c(n) &=& f(k_n),~~~~~~~~\forall n\in \mathbb{N}.
\end{eqnarray}

In words, the above theorem says: ($\Rightarrow$) if for all $x$ the function $f_{c}(x)$ on the circle can be obtained in terms of the function $f(x)$ on the real line by adding contributions from the points $x$, $x \pm l_c$, $x \pm 2 l_c$, etc, then the Fourier transform $f_c(n)$ on the circle can be obtained from the Fourier transform $f(k)$ on the line evaluated at momentum $k=k_n$. The theorem further establishes that the converse ($\Leftarrow$) also holds. Namely, if the Fourier transform $f_c(n)$ on the circle samples the Fourier transform $f(k)$ on the line at momentum $k=k_n$, then the function $f_c(x)$ on the circle can be expressed as the above sum of contributions of the function $f(x)$ on the real line. 

\vspace{0.5cm}

\textbf{Proof:} For ($\Rightarrow$), we have
\begin{eqnarray} \label{eq:imag1}
f_{c}(n) &=& \int_{0}^{l_c} dx~e^{-ik_{n}x} f_{c}(x)~~~\\
&=& \int_{0}^{l_c} dx~e^{-ik_{n}x} \sum_{n\in \mathbb{Z}} f(x+nl_c) \\
&=& \int_{0}^{l_c} dx~e^{-ik_{n}x} \sum_{n\in \mathbb{Z}} ~\int_{\mathbb{R}}\frac{dk}{2\pi} e^{ik(x+nl_c)} f(k) \\
&=& \int_{0}^{l_c} dx~e^{-ik_{n}x}  \int_{\mathbb{R}}\frac{dk}{2\pi} e^{ikx} f(k) \sum_{n\in \mathbb{Z}}e^{iknl_c}~~~\\
&=&\frac{2\pi}{l_c} \int_{0}^{l_c} dx~e^{-ik_{n}x}  \int_{\mathbb{R}}\frac{dk}{2\pi} e^{ikx}f(k) \sum_{m\in \mathbb{Z}} \delta(k-k_{m})~~~\nonumber\\
&=& \frac{1}{l_c} \int_{0}^{l_c} dx~e^{-ik_{n}x}  \sum_{m\in \mathbb{Z}}~ e^{ik_{m}x} f(k_{m})\\
&=& \sum_{m\in \mathbb{Z}} f\left(k_{m}\right) \frac{1}{l_c} \int_{0}^{l_c}dx~ e^{i(k_{m}-k_{n})x}\\
&=& \sum_{m\in \mathbb{Z}} f\left(k_{m}\right) \delta_{m,n} \\
&=& f\left(k_{n}\right). \label{eq:imag2}
\end{eqnarray}
Here we used that
\begin{equation}
\sum_{n \in \mathbb{Z}} e^{ian} = 2\pi \sum_{m\in \mathbb{Z}} \delta(a-2\pi m),    
\end{equation}
which implies
\begin{eqnarray}
\sum_{n\in \mathbb{Z}} e^{iknl_c} &=& 2\pi \sum_{m \in \mathbb{Z}}\delta(kl_c-2\pi m) \\&=& \frac{2\pi}{l_c} \sum_{m \in \mathbb{Z}}\delta(k-\frac{2\pi}{l_c}m) \\
    &=& \frac{2\pi}{l_c} \sum_{m \in \mathbb{Z}}\delta(k-k_{m}). 
\end{eqnarray}
For the converse ($\Leftarrow$) we have
\begin{eqnarray}
 f_c(x) &=& \frac{1}{l_c} \sum_{n\in \mathbb{Z}} e^{ik_nx} f_c(n) \\
 &=&  \frac{1}{l_c} \sum_{n \in \mathbb{Z}} e^{i k_n x} f(k_n) \\
 &=&  \frac{1}{l_c} \sum_{n \in \mathbb{Z}} e^{i k_n x} \int_{\mathbb{R}} dy~e^{-ik_ny}f(y)\\
 &=& \int_{\mathbb{R}} dy~f(y)~\frac{1}{l_c} \sum_{n \in \mathbb{Z}} e^{i k_n (x-y)} \\
 &=& \int_{\mathbb{R}} dy~f(y)~ \sum_{n\in \mathbb{Z}} \delta(x-y+nl_c) \\
 &=& \sum_{n\in \mathbb{Z}} f(x+nl_c),
\end{eqnarray}
where we used that 
\begin{eqnarray}
 \sum_{n\in\mathbb{Z}} e^{ik_n x} &=&  \sum_{n\in\mathbb{Z}} e^{i\frac{2\pi}{l_c}xn} \\
 &=& 2\pi \sum_{n\in \mathbb{Z}} \delta\left(x \frac{2\pi}{l_c}-2\pi n\right)~~~~~~\\
 &=& l_c \sum_{n\in \mathbb{Z}} \delta(x - n l_c).
\end{eqnarray}
This completes the proof of the above theorem. Next we apply it to several cases of interest in the current work.

\vspace{0.5cm}
\begin{center}
\textit{B. Entangler}
\end{center}
\vspace{0.5cm}

Given an entangling profile $g(x)$ on the line, we define an entangling profile on the circle $g_c(x)$ as
\begin{equation}
    g_c(x) = \sum_{n\in \mathbb{Z}} g(x + n l_c).
\end{equation}
It follows from the above theorem that the Fourier transforms $g(k)$ and $g_c(n)$ on the line and on the circle
\begin{eqnarray}
 g(k) &\equiv& \int_{\mathbb{R}} dx~e^{-ikx}g(x),\\
 g_c(n) &\equiv& \int_{0}^{l_c} dx~e^{-ik_nx}g_{c}(x),
\end{eqnarray}
are related through
\begin{equation} \label{eq:gcgkn}
    g_c(n) = g(k_n).
\end{equation}


\vspace{0.5cm}
\begin{center}
\textit{C. Two-point correlators}
\end{center}
\vspace{0.5cm}

Next we consider a state $\ket{\Psi}$ on the line characterized by annihilation operators $b(k)$
as
\begin{eqnarray}
&& b(k) \ket{\Psi} = 0,~~~~~ \forall k\in \mathbb{R},\\ 
&& b(k) \equiv \sqrt{\frac{\beta(k)}{2}}\phi(k) + \frac{i}{\sqrt{2\beta(k)}}\pi(k). \label{eq:bk3}
\end{eqnarray}
Its correlation functions in momentum space read 
\begin{eqnarray}
 \bra{\Psi} \phi(k)\phi(k')\ket{\Psi} &\equiv& \delta(k+k') ~ C_{\phi\phi}(k),\\
 \bra{\Psi} \pi(k)\pi(k')\ket{\Psi} &\equiv& \delta(k+k') ~ C_{\pi\pi}(k),
\end{eqnarray}
for
\begin{equation}
    C_{\phi\phi}(k) = \frac{1}{2\beta(k)},~~~~~C_{\pi\pi}(k) =  \frac{\beta(k)}{2},
\end{equation}
whereas in real space we have
\begin{eqnarray}  
C_{\phi\phi}(x) &\equiv& \bra{\Psi} \phi(x)\phi(0)\ket{\Psi} = \int_{\mathbb{R}} \frac{dk}{2\pi} ~  e^{i kx}C_{\phi\phi}(k),~~~\\
C_{\pi\pi}(x) &\equiv& \bra{\Psi} \pi(x)\pi(0)\ket{\Psi} = \int_{\mathbb{R}} \frac{dk}{2\pi} ~  e^{i kx}C_{\pi\pi}(k).~~~~~ 
\end{eqnarray}

Similarly, we consider a state $\ket{\Psi_c}$ on the circle characterized by annihilation operators $b(n)$
as
\begin{eqnarray}
&& b(n) \ket{\Psi_c} = 0,~~~~~ \forall n\in \mathbb{Z},\\ 
&& b(n) \equiv \sqrt{\frac{\beta_c(n)}{2}}\phi(n) + \frac{i}{\sqrt{2\beta_c(n)}}\pi(n), \label{eq:bn3}
\end{eqnarray}
which has correlators 
\begin{eqnarray}
 \bra{\Psi} \phi(n)\phi(n')\ket{\Psi} &=& \delta_{n,-n'} C_{c,\phi\phi}(n),\\
 \bra{\Psi} \pi(n)\pi(n')\ket{\Psi} &=& \delta_{n,-n'} C_{c,\pi\pi}(n),
\end{eqnarray}
for
\begin{equation}
    C_{c,\phi\phi}(n) = \frac{1}{2\beta_c(n)},~~~~~C_{c,\pi\pi}(n) =  \frac{\beta_c(n)}{2},
\end{equation}
or, in real space,
\begin{eqnarray}  
C_{c,\phi\phi}( x) &\equiv& \bra{\Psi_c} \phi(x)\phi(0)\ket{\Psi_c} \\
&=&  \frac{1}{l_c}\sum_{n\in \mathbb{Z}} e^{ik_nx} C_{c,\phi\phi}(n), ~~~~\\
C_{c,\pi\pi}(x) &\equiv& \bra{\Psi_c} \phi(x)\phi(0)\ket{\Psi_c} \\
&=& \frac{1}{l_c} \sum_{n \in \mathbb{Z}} e^{i k_nx} C_{c,\pi\pi}(n).~~~~
\end{eqnarray}

Above we did not yet assume any relation between the line annihilation operators $b(k)$ in Eq. \eqref{eq:bk3} and the circle annihilation operators $b(n)$ in Eq. \eqref{eq:bn3}. Let us now assume that they are related through
\begin{equation} \label{eq:betanbetak}
    \beta_c(n) = \beta(k_n).
\end{equation}
(As reviewed below, this relation naturally holds for ground states of local Hamiltonians. It also holds, by construction, in the cMERA). It follows that the momentum space correlators on the line and on the circle are related by
\begin{equation}
     C_{c,\phi\phi}(n) = C_{\phi\phi}(k_n),~~~~ C_{c,\pi\pi}(n) = C_{\pi\pi}(k_n).~~~~ 
\end{equation}
Then the above theorem implies that the real space correlators on the line and on the circle are related by
\begin{eqnarray}
 C_{c,\phi\phi}(x) &=& \sum_{n\in \mathbb{Z}} C_{\phi\phi}(x+n l_c), \nonumber\\ 
 C_{c,\pi\pi}(x) &=& \sum_{n\in \mathbb{Z}} C_{\pi\pi}(x+n l_c).  \label{eq:correlators_line_circle}
\end{eqnarray}

\vspace{0.5cm}
\begin{center}
\textit{D. Ground states of local Hamiltonians}
\end{center}
\vspace{0.5cm}

Let us specialize the above discussion to the ground state of a generic non-interacting local QFT Hamiltonian. Here we follow closely the analysis of Y. Zou in an appendix of \cite{Yijian} on the line, which we (trivially) extend to include also the circle. Let $H^{\QFT}$ denote a Hamiltonian of the form
\begin{eqnarray}
 &&H^{\QFT} \equiv \frac{1}{2}\int_{\mathbb{R}} dx\sum_{l} \left( a_l \left(\partial_x^{l} \pi(x)\right)^2 + b_l \left( \partial_x^{l} \phi(x)\right)^2 \right) ~~~~~~~~\\
 &&= \frac{1}{2}\int_{\mathbb{R}} dk \sum_{l}  k^{2l}\left( a_l \pi(-k)\pi(k) + b_l \phi(-k) \phi(k) \right) ~~~~ \\
 &&= \int_{\mathbb{R}} dk~ E(k) b^{\dagger}(k) b(k),
\end{eqnarray}
where $E(k)$ is the single particle energy $E(k)$ and $b(k)$ is an annihilation operator, 
\begin{equation}
    b(k) \equiv \sqrt{\frac{\beta(k)}{2}} \phi(k) + \frac{i}{\sqrt{2\beta(k)}} \pi(k).
\end{equation}
Let us first determine $E(k)$ and $\beta(k)$. Since 
\begin{eqnarray}
&& b^{\dagger}(k) b(k) + b^{\dagger}(-k) b(-k) =~~~~~~~~~~~~~~~~~~~~~ \\
&&~~~~~~~~~~~~~~~~~ \beta(k) \phi(-k)\phi(k) + \frac{1}{\beta(k)} \pi(-k) \pi(k),~~~~~~ \nonumber
\end{eqnarray}
from requiring
\begin{eqnarray}
&& E(k) \left(b^{\dagger}(k) b(k) + b^{\dagger}(-k) b(-k)\right)  =~~~~~~~~~~~\\
&&~~~~~~~\sum_l k^{2l}\left( a_l \pi(-k)\pi(k) + b_l \phi(-k) \phi(k) \right),~~\nonumber
\end{eqnarray}
we conclude that 
\begin{equation}
     E(k) \beta(k) = Q(k),~~ E(k) \frac{1}{\beta(k)} = P(k),~~~
\end{equation}
for $P(k) \equiv \sum_l a_l k^{2l}$ and $Q(k) \equiv \sum_l b_l k^{2l}$, or
\begin{equation}
     E(k) = \sqrt{Q(k)P(k) },~~~\beta(k) = \sqrt{\frac{Q(k)}{P(k)}}.
\end{equation}

Let us repeat this calculation for a Hamiltonian $H^{\QFT}_c$ obtained from the same Hamiltonian density integrated on the circle, namely
\begin{eqnarray}
 &&H^{\QFT}_c \\
 &\equiv& \frac{1}{2}\int_{0}^{l_c} \!\! dx\sum_{l} \left( a_l \left(\partial_x^{l} \pi(x)\right)^2 + b_l \left( \partial_x^{l} \phi(x)\right)^2 \right) ~~~~\\
 &=& \frac{1}{2} \sum_n \sum_{l}  (k_n)^{2l}\left( a_l \pi(-n)\pi(n) + b_l \phi(-n) \phi(n) \right) ~~~~ \\
 &=& \sum_n~ E_c(n) b^{\dagger}(n) b(n),
\end{eqnarray}
where
\begin{equation}
    b(n) \equiv \sqrt{\frac{\beta_c(n)}{2}} \phi(n) + \frac{i}{\sqrt{2\beta_c(n)}} \pi(n).
\end{equation}
We have 
\begin{eqnarray}
&& b^{\dagger}(n) b(n) + b^{\dagger}(-n) b(-n) =~~~~~~~~~~~~~~~~~~~~~ \\
&&~~~~~~~~~~~~~~~~~ \beta_c(n) \phi(-n)\phi(n) + \frac{1}{\beta_c(n)} \pi(-n) \pi(n),~~~~~~ \nonumber
\end{eqnarray}
and, arguing similarly as above we arrive at 
\begin{equation}
     E_c(n) \beta_c(n) = Q_c(n),~~ E_c(n) \frac{1}{\beta_c(n)} = P_c(n)~~~
\end{equation}
for $P_c(n) \equiv \sum_l a_l (k_n)^{2l}$ and $Q_c(l) \equiv \sum_l b_l (k_n)^{2l}$, or
\begin{equation}
     E_c(n) = \sqrt{Q_c(n)P_c(n) },~~~\beta_c(n) = \sqrt{\frac{Q_c(n)}{P_c(n)}}.
\end{equation}

Finally, we notice that since $Q_c(n) = Q(k_n)$ and $P_c(n) = P(k_n)$, it follows that $\beta(k)$ on the line and $\beta_c(n)$ on the circle are related by $\beta_c(n) = \beta(k_n)$, that is Eq. \eqref{eq:betanbetak}. Therefore the ground states $\ket{\Psi^{\QFT}}$ on the line and $\ket{\Psi_c^{\QFT}}$ on the circle, given by
\begin{eqnarray}
 b(k) \ket{\Psi^{\QFT}} &=& 0,~~~~~~\forall k \in \mathbb{R},\\
 b(n) \ket{\Psi^{\QFT}_c} &=& 0,~~~~~~\forall n \in \mathbb{Z},
\end{eqnarray}
have correlators that are indeed related by Eq. \eqref{eq:correlators_line_circle},
\begin{eqnarray}
 C^{\QFT}_{c,\phi\phi}(x) &=& \sum_{n\in \mathbb{Z}} C^{\QFT}_{\phi\phi}(x+n l_c), \nonumber\\ 
 C^{\QFT}_{c,\pi\pi}(x) &=& \sum_{n\in \mathbb{Z}} C^{\QFT}_{\pi\pi}(x+n l_c).  \label{eq:correlators_line_circle_QFT}
\end{eqnarray}

\vspace{0.5cm}
\begin{center}
\textit{E. cMERA}
\end{center}
\vspace{0.5cm}

On the line, consider an entangler of the form
\begin{equation}
 \tilde{K}(s) = \frac{1}{2}\int_{\mathbb{R}} dk\, \tilde{g}(s,k) [\pi(-k)\phi(k) + \phi(-k)\pi(k)].   
\end{equation}
The cMERA state $\ket{\Psi^{\Lambda}(s)}$, defined by
\begin{equation}
    \ket{\Psi^{\Lambda}(s)} \equiv \mathcal{P}\exp \left(-i \int_0^{s} ds' \tilde{K}(s')  \right) \ket{\Lambda},
\end{equation}
is then annihilated by operators $\tilde{b}^{\Lambda}(s,k)$,
\begin{equation}
    \tilde{b}^{\Lambda}(s,k) \ket{\Psi^{\Lambda}(s)} = 0, ~~~\forall k \in \mathbb{R},
\end{equation}
such that
\begin{equation} \label{eq:ansatz_b}
    \tilde{b}^{\Lambda}(s,k) \equiv \sqrt{\frac{\tilde{\beta}(s,k)}{2}}\phi(k) + \frac{i}{\sqrt{2\tilde{\beta}(s,k)}}\pi(k).
\end{equation}
We saw that $\tilde{\beta}(s,k)$ obeys the differential equation
\begin{equation} \label{eq:diff1}
    \frac{\partial}{\partial s} \tilde{\beta}(s,k) = -2\tilde{g}(s,k) \tilde{\beta}(s,k),~~~~\forall k \in \mathbb{R},
\end{equation}
with initial condition
\begin{equation} \label{eq:initial_R}
    \tilde{\beta}(s=0,k) = \Lambda,~~~~\forall k \in \mathbb{R}.
\end{equation}

On the circle, consider now an entangler of the form
\begin{equation}
 \tilde{K}_c(s) = \frac{1}{2} \sum_{n\in \mathbb{N}} \tilde{g}_c(s,n) [\pi(-n)\phi(n) + \phi(-n)\pi(n)],
\end{equation}
The cMERA state $\ket{\Psi^{\Lambda}(s)}$
\begin{equation}
    \ket{\Psi^{\Lambda}_c(s)} \equiv \mathcal{P}\exp \left(-i \int_0^{s} ds' \tilde{K}_c(s')  \right) \ket{\Lambda_c}
\end{equation}
is annihilated by operators $b^{\Lambda}(s,n)$,
\begin{eqnarray}
 &&\tilde{b}^{\Lambda}(s,n) \ket{\Psi^{\Lambda}_c(s)} = 0, ~~~\forall n \in \mathbb{Z},\\
 &&\tilde{b}^{\Lambda}(s,n) \equiv \sqrt{\frac{\tilde{\beta}_c(s,n)}{2}}\phi(n) + \frac{i}{\sqrt{2\tilde{\beta}_c(s,k)}}\pi(n),
\end{eqnarray}
where $\beta_c(s,n)$ obeys the differential equation
\begin{equation} \label{eq:diff3}
    \frac{\partial}{\partial s} \tilde{\beta}_c(s,n) = -2\tilde{g}_c(s,n) \tilde{\beta}_c(s,n),~~~~\forall n \in \mathbb{Z},
\end{equation}
with the initial condition
\begin{equation} \label{eq:initial_c}
    \tilde{\beta}_c(s=0,n) = \Lambda,~~~~\forall n \in \mathbb{Z}.
\end{equation}

Clearly, Eqs. \eqref{eq:diff1}-\eqref{eq:initial_R} for $\tilde{\beta}(s,k)$ on the line and Eqs.  \eqref{eq:diff3}-\eqref{eq:initial_c} for $\tilde{\beta}_c(s,n)$ on the circle are very similar. If, following the proposal of this work (see Eq. \eqref{eq:gc} in the main text), we set the entangling profile on the line and on the circle to be related by Eq. \eqref{eq:gcgkn}, that is
\begin{equation} \label{eq:gcgkn3}
    \tilde{g}_c(s,n) = \tilde{g}(s,k_n).
\end{equation}
then Eqs. \eqref{eq:diff1}-\eqref{eq:initial_R} and Eqs. \eqref{eq:diff3}-\eqref{eq:initial_c} imply that also 
\begin{equation}
    \tilde{\beta}_c(s,n) = \tilde{\beta}(s,k_n),~~~~~\forall n \in \mathbb{Z}.
\end{equation}
This last equation is analogous to condition \eqref{eq:betanbetak}.
Therefore the correlators of the cMERA state on the line and on the circle are indeed related by Eq. \eqref{eq:correlators_line_circle}. That is,
\begin{eqnarray}
 C^{\Lambda}_{c,\phi\phi}(s,x) &=& \sum_{n\in \mathbb{Z}} C^{\Lambda}_{\phi\phi}(s,x+n l_c), \nonumber\\ 
 C^{\Lambda}_{c,\pi\pi}(s,x) &=& \sum_{n\in \mathbb{Z}} C^{\Lambda}_{\pi\pi}(s, x+n l_c).  \label{eq:correlators_line_circle_cMERA}
\end{eqnarray}

\vspace{0.5cm}
\textbf{Appendix A4: Generalizations beyond the circle}
\vspace{0.5cm}

In this last appendix we briefly comment on possible generalizations of the results presented in this paper, based on applying the method of images to other geometries.

In its essence, the method of images is based on generating solutions of a linear differential equation (the equations of motion) by summing over different solutions related to each other by a discrete symmetry transformation (such as a translation in the line by a distance $l_c$, which allows us to turn the line into a circle of size $l_c$). This is possible for linear systems, and thus free theories. The same techniques thus admit immediate generalizations. 

\vspace{0.5cm}
\begin{center}
\textit{A. Higher dimensions}
\end{center}
\vspace{0.5cm}


One generalization is to obtain the cMERA on an $N$- dimensional torus $T^N$ from $\mathbb{R}^N$, since an $N$-torus can be thought of as an orbifold of $\mathbb{R}^N$.

This is done by repeating the computation, with a sum over images in $N$ directions separately. 
To be precise, on an $N$ torus, a field $\phi(x)$ is expressible as
\be
\phi(x) = \sum_{\{n_i\}} \left(\prod_{i}^N  \frac{1}{\sqrt{l_{c_i}}}\right) e^{i k_{\{n_i\}}.x} \phi(\{n_{i}\}), 
\ee
 where $x$ is an $N$ component vector, and $k_{\{n_i\}}$ also an $N$ component vector given by
 \be
 k_{\{n_i\}} =\{ \frac{2\pi n_i}{l_{c_i}}  \}, \qquad n_i \in \mathbb{Z},
\ee
and $l_{c_i}$ is the periodicity of each cycle $i$ in $T^N$. A similar epxression applies to $\pi(x)$.
The cMERA entangler would take the form
\begin{align}
&K_{T^N}(s) \\
&=\frac{1}{2}  \int_{T^N} d^Nx d^N y  g_{T^N}(s, x -y) [\pi(x) \phi(y) + \phi(x)\pi(y)] \\
&=\frac{1}{2}\sum_{\{n_i \in \mathbb{N}\}} g_{T^N}(s, \{n_i\}) [\pi(\{-n_i\}) \phi(\{ n_i  \})  \nonumber\\
&+ \pi(\{-n_i\}) \phi(\{ n_i  \})].
\end{align}
Here $g_{T^N}(s,\{n_i\})$ again correspond to the Fourier transform of $g_{T^N}(s, x -y)$
The method of images would imply that one can generate an admissible entangler on $T^N$ using the entangler on the $\mathbb{R}^N$. i.e. 
\be
g_{T^N}(s,x) = \sum_{\{n_i\}} g_{\mathbb{R}^N}(s, x_{\{n_i\}}), \qquad x_{\{n_i\}} \equiv \{x_i + n_i l_{c_i}\}.
\ee

On $T^2$ for example, this procedure generates doubly periodic functions. 
The method of images is widely applied in CFT2 to generate correlation functions on $T^2$ in minimal models, which admit free field representations.

\vspace{0.5cm}
\begin{center}
\textit{B. Open boundaries}
\end{center}
\vspace{0.5cm}

 
One can also generate systems with boundaries using the method of images.  
Consider for concreteness a 1-dimensional quantum system on the half-line. i.e. There is a boundary at $x=0$.
The effect of the boundary can be treated as a mirror, in which inserting an operator $\mathcal{O}(x,t)$  in the presence of the boundary is equivalent to inserting a pair of operators $\mathcal{O}(x,t)$ and $\mathcal{O}(-x,t)$ in the theory which is defined on the real line with no boundaries. In complex coordinates $z = t+ i x$ and $\bar z = t- i x$, this gives 
\be \label{eq:folding}
\langle \mathcal{O}(z, \bar z) \cdots \rangle_{UHP} = \langle \mathcal{O}(z) \bar{\mathcal{O}}^P(z*) .... \rangle_{\mathbb{R}^2},
\ee

where $\mathcal{O}^P(z*)$ corresponds to the parity transformed anti-holomorphic part of the operator $\mathcal{O}(z,\bar z)$. The parity transformation turns the anti-holomorphic operator into a holomorphic one, now inserted at $z*$. As an example of such a parity transformation,  considers a free bosonic field $\phi(z)$. Then
\be
(\partial_{\bar z} \bar \phi)^P ( z*) \equiv \eta_P \partial_z \phi(z*),
\ee
where $\eta_P = \pm1$, depending on the precise boundary condition that is imposed.
For example the Dirichlet boundary condition that sets $\phi(z,\bar z) =0$ at the boundary located on the real line gives $\eta_P = 1$, whereas the Neumann
boundary condition that sets the derivative of $\phi$ to zero gives $\eta_P = -1$. 
We note that in this context of conformal boundary conditions, the method of images encapsulated in (\ref{eq:folding}) works beyond free field theories because in two dimensional CFTs holomorphic and anti-holomorphic conformal transformations are independent.

This fact thus allows one to generate cMERA entanglers applicable to theories with boundaries from entanglers on the real line using the method of images. 
Consider the specific case of a free boson $\phi(x)$ again. 

Suppose we impose Neumann boundary condition at $x=0$
\be
\partial_x\phi(0) = 0.
\ee
The mode expansion of $\phi_{\mathbb{R}>0}(x)$  and $\pi_{\mathbb{R}>0}(x)$
would then be given by
\begin{align}
&\phi_{\mathbb{R}>0}(x) = \frac{2}{\sqrt{2\pi}}\int_0^
\infty dk  \cos(k x) \phi_{\mathbb{R}>0}(k), \\
& \pi_{\mathbb{R}>0}(x) = \frac{2}{\sqrt{2\pi}}\int_0^
\infty dk  \cos(k x) \pi_{\mathbb{R}>0}(k).
\end{align}
We note that this implies 
\be \label{eq:reflect}
\phi_{\mathbb{R}>0}(x) = \frac{1}{2}(\phi_{\mathbb{R}} (x) + \phi_{\mathbb{R}>0}(-x)),
\ee
and that 
\be
\phi_{\mathbb{R}>0}(k) = \frac{1}{2}(\phi_{\mathbb{R}}(k) + \phi_{\mathbb{R}}(-k))
\ee
and similarly for $\pi_{\mathbb{R}>0}(x)$ and its Fourier modes.
In other words, we have selected a boundary condition that imposes that the Fourier modes are even in $k$.

Consider the entangler characterized by the function $g(s, x)$ discussed in the main text.  On the half-line is given by
\begin{align}
&\tilde K_{\mathbb{R}>0}(s) = \frac{1}{2} \int_{\mathbb{R}^2>0} dx dy  \, \tilde g_{\mathbb{R}>0}(x,y) \bigg(\phi_{\mathbb{R}>0}(x) \pi_{\mathbb{R}>0}(y)  \nonumber \\
&+ \pi_{\mathbb{R}>0}(x) \phi_{\mathbb{R}>0}(y)\bigg).
\end{align}

Now using (\ref{eq:reflect}), we notice that this can be unfolded into an integral over $\mathbb{R}^2$
\be
\tilde K_{\mathbb{R}>0}(s) =  \frac{1}{2} \int_{\mathbb{R}^2} dx dy \, \tilde g(x,y)  \left( \phi(x) \pi(y) + \pi(x) \phi(y)\right),
\ee
if we define 
\be
\tilde g(-|x|,|y|) = \tilde g(|x|,-|y|) = \tilde g(-|x|,-|y|) = \tilde g(|x|, |y|)
\ee
This implies that the desired solution of an entangler on the half line can be generated from that the entangler on the real line $\tilde g_{\mathbb{R}}(x-y)$ by
\begin{align}
&\tilde g_{\mathbb{R}>0}(x,y) = \frac{1}{4} (\tilde g_{\mathbb{R}}(x-y) +  \tilde g_{\mathbb{R}}(-x-y) \nonumber \\
&+  \tilde g_{\mathbb{R}}(x+y) +  \tilde g_{\mathbb{R}}(-x-y)).
\end{align}

 In general beyond free field theories, the method of images are not expected to work. We note however that it is known that in large N theories, such as CFTs with an AdS holographic dual, correlation functions at finite temperatures can be obtained from those at zero temperatures by the method of images -- at least for sufficiently low temperatures.
 Therefore, there are hopes that the method of images that map cMERA to other cMERA defined on a compact space can be applied even if the theory concerned is interacting.


\begin{thebibliography}{99}
\bibitem{Kconstant} Ref. \cite{cMERA} introduced a more general class of scale-dependent entangler $K(s)$, which nevertheless maintained a constant entangling scale $1/\Lambda$. For simplicity, in this work we restrict our attention to scale-independent entanglers $K \not = K(s)$ (which then become scale-dependent entanglers $\tilde{K}(s) \equiv e^{isD} (L+K-D) e^{-isD}$ in the re-scaled picture used in this work). A generalization of our results to a scale-dependent entangler $K(s)$ requires more complex notation but is otherwise straightforward.
\bibitem{cutoff} In the usual picture of cMERA, the UV and IR lengths are given by $1/\Lambda$ and $e^{s}/\Lambda$, respectively. In the re-scaled picture used in this paper, the UV and IR lengths are instead given by $e^{-s}/\Lambda$ and $\Lambda$.
\end{thebibliography}
\end{document}